\documentclass[notitlepage,aps,pra,twocolumn,groupaddress,10pt]{revtex4-1}
\usepackage{stmaryrd}% authblk
\usepackage{amssymb,amsmath,amsthm,amsfonts,amsbsy}
\usepackage{bm,bibunits,color,chngcntr,epsfig,epstopdf,graphicx,dsfont}
\usepackage{hyperref,lipsum,,makecell,mathrsfs,rotating}
\usepackage[english]{babel}
\usepackage[normalem]{ulem}

\newcommand{\be}{\begin{equation}}
\newcommand{\ee}{\end{equation}}
\newcommand{\bea}{\begin{eqnarray}}
\newcommand{\eea}{\end{eqnarray}}
\newcommand{\ket}{\rangle}
\newcommand{\bra}{\langle}

\newcommand{\I}{\mathds{1}}
\newcommand{\ra}{\rightarrow}

\def\C#1{\mathcal #1}

\definecolor{gray}{gray}{0.9}

\usepackage{setspace}
\begin{document}
\newtheorem{theorem}{Theorem}
\newtheorem{prop}[theorem]{Proposition}
\newtheorem{corollary}[theorem]{Corollary}
\newtheorem{open problem}[theorem]{Open Problem}
\newtheorem{conjecture}[theorem]{Conjecture}
\newtheorem{definition}{Definition}
\newtheorem{remark}{Remark}
\newtheorem{example}{Example}
\newtheorem{task}{Task}

\title{General channel capacities from quantum channel-state duality}

\author{Yuan-Dong Liu}
\affiliation{CAS Key Laboratory of Theoretical Physics, Institute of Theoretical Physics,
Chinese Academy of Sciences, Beijing 100190, China \\
School of Physical Sciences, University of Chinese Academy of Sciences, Beijing 100049, China}
\author{Dong-Sheng Wang}
\email{wds@itp.ac.cn}
\affiliation{CAS Key Laboratory of Theoretical Physics, Institute of Theoretical Physics,
Chinese Academy of Sciences, Beijing 100190, China \\
School of Physical Sciences, University of Chinese Academy of Sciences, Beijing 100049, China}

\date{\today}
\begin{abstract}
The quantum channel-state duality permits the characterization of a quantum process through a quantum state, 
referred to as a Choi state. 
This characteristic serves as the impetus for the quantum computing paradigm that utilizes Choi states as information sources.
In this work, 
the fundamental theorems regarding quantum channel capacity are proven when Choi states are considered as sources. 
This achievement enriches the set of capacities associated with quantum channels. 
Moreover, it gives rise to novel opportunities for the comprehension of superadditivity phenomena 
and the discovery of new classes of quantum error-correction codes.
\end{abstract}

\maketitle

\begin{spacing}{1.2}

\section{Introduction}

Quantum error correction is essential for quantum information processing~\cite{NC00,LB13}.
The quantum channel capacity theorems, 
as the quantum version of Shannon's seminal information theory,
guide the design of good quantum error-correction codes
given an error model described as a noise channel.
A channel capacity, as the upper bound of encoding rates 
over all possible codes,
is extremely hard to obtain  
due to the notable superadditivity phenomena 
of quantum channels~\cite{Wil17,Wat18}.

Given a quantum channel, its capacity is not unique
and many types of capacities have been established using distinct methods~\cite{Wil17,Wat18}.
The nonadditivity of capacity measures, 
such as coherent information~\cite{SN96,Llo97,BNS98,SY08} and Holevo mutual information~\cite{Hol99,Has09}, 
and also the private capacity~\cite{Dev05,LLS+14},
poses great challenges for the proof of capacity theorems
and design of good codes.
Efforts have been made to systematically understand capacities and more general quantum protocols~\cite{DHW04,ADHW09,CLS17}.
However, it is unclear whether there could be more capacities or not.
%In recent years, 
%diverse approaches have been developed to 
%understand channel capacities~\cite{LLS18,LLS+23,WLWL24}. 

%This motivates the scheme of using Choi states as information carrier.
%A fundamental question is how to define quantum channel capacities for Choi states.

% Choi channel-state duality converts dynamics into state, 
% it leads to QvN, 
% and here they motivate our model of coding using quantum control unit and 
% quantum memory unit. 
% We call this ``quantum vN coding''

A distinct feature of quantum physics is
known as the quantum channel-state duality~\cite{Jam72,Cho75}, 
which can treat a quantum process as a quantum state,
usually called a Choi state.
It has led to novel development such as
quantum process tomography~\cite{NC00},
quantum superchannel and comb theory~\cite{CDP08},
and more recently, the dynamical resource theory~\cite{CG19}
and quantum von Neumann architecture which manipulates 
information stored as Choi states~\cite{W22_qvn,W24_qvn}.
From the perspective of capacity theory, 
a fundamental question then emerges: 
is it feasible to define quantum channel capacities for Choi states?

In this work, we answer this question in affirmative. 
We focus on quantum capacities of a quantum channel,
and add to the study of channel capacity 
with a unifying understanding and proof of quantum capacities,
which we believe can also be extended to the classical capacities.
With this framework, 
we prove two capacity theorems for Choi states,
which are the analog of the usual quantum capacity~\cite{BKN00,HHW+08,Kle07}
and entanglement-assisted quantum capacity~\cite{BSS+99,BSST02,BDH+14}.
This not only enriches the family of capacities,
but also enables fault-tolerant quantum von Neumann architecture
including quantum units for storage, control,
computing and communication.
The Choi coding models also show different features of nonadditivity,
and shed light on the design of new types of quantum codes.

\section{Standard Coding}

% \begin{figure}[h!]
%     \centering
%     \includegraphics[width=0.4\textwidth]{code}
%     \caption{A schematic of quantum coding with a 
%     superchannel $\hat{\C S}$ serving as the coding operation converting 
%     $n$ uses of a channel $\Phi$ into $k$ uses
%     which approximates the identity channel.}
%     \label{fig:code}
% \end{figure}

Quantum channels are completely positive trace-preserving maps, 
and from dilation a quantum channel $\Phi$ can be realized as 
\be \Phi(\rho)= \text{tr}_a V \rho V^\dagger, \ee 
for input system states $\rho$ and an isometry $V$,
which can be expressed as $V=U|0\ket=\sum_i K_i |i\ket$
for a unitary operator $U$ acting on the system 
and $K_i$ are known as Kraus operators~\cite{NC00}.
The trace is over an ancilla with $\{|i\ket\}$ as an orthonormal basis,
while if the system is traced out
one obtains the complementary channel $\Phi^c$ of $\Phi$.

% \be \Phi(\rho)= \text{tr}_a \C U (\rho\otimes |0\ket \bra 0|)=\sum_i K_i \rho K_i^\dagger, \ee
% for $\rho\in \C D(\C H_1)$, 
% $\C U$ as the superoperator form of a unitary operator, 
% the trace $\text{tr}_a$ is over an ancilla a at an initial state $|0\ket$,  
% which realizes Kraus operators as $K_i=\bra i|U|0\ket$, 
% with $\{|i\ket\}$ as an orthonormal basis of the ancilla. 
% In the following, we consider dimension-preserving channels for convenience.

Given a noise channel $\Phi$,
a coding protocol in general involves the conversion of $n$ parallel uses of $\Phi$
into $k$ approximate uses of an identity channel, for positive integers $n$ and $k\leq n$.
The value $\alpha={k}/{n}$ is known as the coding rate,
and the supremum of all achievable rates is known as 
the quantum capacity of the channel for such a coding protocol. %~\cite{Wat18}. 
Two standard quantum capacities are the quantum capacity without assistance 
\be Q(\Phi)= I_{c_r}(\Phi) \label{eq:capacityq}\ee 
for $I_{c_r}(\Phi):=\lim_{n\ra \infty}\frac{1}{n} \max_\rho I_c(\rho,\Phi^{\otimes n})$ 
denoting the regularized coherent information, and 
$I_c(\rho,\Phi):=S(\Phi(\rho))-S(\Phi^c(\rho))$ is the usual coherent information
for an input source state $\rho$ over a channel $\Phi$,
and the entanglement-assisted (EA) quantum capacity
\be Q_{EA}(\Phi)=\max_\rho \frac{1}{2} I(\rho,\Phi) \label{eq:capacityea}\ee 
for $I(\rho,\Phi):=S(\rho)+I_c(\rho,\Phi)$ as the quantum mutual information,
and $S(\rho)$ as the von Neumann entropy of the state $\rho$.
A notable fact is that while $I(\rho,\Phi)$ is subadditive~\cite{AC97}, 
$I_c(\rho,\Phi)$ is not, which leads to the necessity of the regularization 
for the expression of $Q(\Phi)$ and difficulty to compute it for an arbitrary channel $\Phi$~\cite{SN96,Llo97,BNS98,SY08}.

% In general, given a type of coding scheme, 
% the quantum capacity of a channel $Q(\Phi)$,
% is defined as the supremum of all achievable rate $\alpha$
% for $k=\lfloor \alpha n \rfloor$ 
% with arbitrarily large $n$ but small $\epsilon$~\cite{Wat18}.

% The quantum capacity is 
% \be Q(\Phi)=\max_\rho I_{c_r}(\rho,\Phi) \ee 
% The EA quantum capacity is 
% \be Q_{EA}(\Phi)=\max_\rho \frac{1}{2} I(\rho,\Phi). \ee 

A general framework that can take the two capacities above as special cases 
is by using the channel-state duality and superchannels~\cite{CDP08,MW14,LM15,WLWL24}.
The channel-state duality represents a channel $\Phi$ as 
\be \omega_{\Phi} := \Phi \otimes \I (\omega), \ee 
usually known as a Choi state,
and the Bell state, also known as an ebit, 
is $|\omega\ket:=\frac{1}{\sqrt{d}} \sum_i |ii\ket$, 
$\omega:=|\omega\ket \bra \omega|$,
for $d$ as the input system dimension.
%$d=\text{dim}(\C H_1)$,
The operations that preserve the form of Choi states are superchannels,
whose actions are 
\be \hat{\C S} (\Phi)(\rho)= \text{tr}_{a} \C V_2  (\Phi \otimes \I) \C V_1 (\rho),  
\label{eq:superchannel} \ee
for a pre- isometry $\C V_1$ that requires an ancilla $a_1$
and a post isometry $\C V_2$ that requires another ancilla $a_2$, and $a=a_1a_2$.
A noise-free quantum memory between the pre- and post operations is in general required 
to realize a superchannel $\hat{\C S}$.
It can also be represented as 
$\hat{\C S}(\omega_{\Phi})=\sum_\mu S_\mu \omega_{\Phi} S_\mu^\dagger$ 
for a set of bipartite Kraus operators $S_\mu$
(See Appendix~\ref{sec:superc} for more details). 
Note we put a hat on the symbols for superchannels and their capacities. 

% A superchannel $\hat{\C S}: \C C(\C H_1 \otimes \C H_2) \ra \C C(\C H_{1'} \otimes \C H_{2'})$ can be realized as
% \be \hat{\C S} (\Phi)(\rho)= \text{tr}_a \C V \; (\Phi \otimes \I)\; \C U (\rho\otimes |0\ket \bra 0|),  \ee
% for $\rho \in \C D(\C H_{1'})$, 
% $\C U$ and $\C V$ are unitary operators,
% and a is an ancilla with an initial state $|0\ket$.

A coding protocol for a noise channel $\Phi$ is a superchannel $\hat{\C S}$ so that 
\be F(\omega^{\otimes k},(\hat{\C S} (\Phi^{\otimes n})\otimes \I^{\otimes k})(\omega^{\otimes k}))
\geq 1-\epsilon, \label{eq:codeerror}\ee 
with $\epsilon\in [0,1]$ and
the state fidelity function $F(\rho,\sigma):=\|\sqrt{\rho}\sqrt{\sigma}\|_1^2$,
with $\|\cdot\|_1$ denoting the trace norm~\cite{WLWL24}.
The fidelity above is the fidelity between ebits $\omega^{\otimes k}$
and the Choi state of $\hat{\C S} (\Phi^{\otimes n})$,
also known as the average entanglement fidelity~\cite{Sch96},
which is a proper measure of the coding accuracy~\cite{Kle07}.
The usual quantum coding is the factorized setting with a pre- encoding isometry
and a post decoding channel, 
and the EA coding is the more generic setting.

%\be F_E(\I^{\otimes k},\hat{\C S} (\Phi^{\otimes n}))\geq 1-\epsilon, \label{eq:codeerror}\ee 

%the average entanglement fidelity~\cite{Sch96} 
%\be F_E(\Phi, \Psi):=  F(\Phi \otimes \I (\omega),  \Psi \otimes \I (\omega)), \ee

%Here the distance $D_\diamond$ is the diamond-norm distance,
%$\epsilon\in [0,1]$ is the accuracy of the code.

%This can be extended to multi-stage case, as a serial of superchannels. 

% \begin{definition}
% (Quantum capacity of a channel~\cite{Wat18}) 
% Let $\Phi\in C(\C X, \C Y)$ be a channel, and an integer $m=\lfloor \alpha n \rfloor$ 
% for all but finitely many positive integers $n$ and an achievable rate $\alpha \geq 0$, 
% there exists channels $\Phi\in C(\C Z^{\otimes m}, \C X^{\otimes n})$ 
% and $\C D\in C(\C Y^{\otimes n}, \C Z^{\otimes m})$ such that 
% \be D_\diamond (\I^{\otimes m}, \C D \Phi^{\otimes n} \Phi) \leq \epsilon\ee
% for every choice of a positive real number $\epsilon$
% and the quantum capacity of $\Phi$, denoted $Q(\Phi)$,
% is defined as the supremum of all $\alpha$.
% \end{definition}

\section{Choi coding}

The formalism above needs to be extended 
when the Choi state $\omega_\Phi$ instead of the channel $\Phi$ is provided.
Namely, when the input sources are Choi states,
distinct coding protocols are possible. 
This is also motivated by the dynamical quantum resource theory~\cite{CG19}
and von Neumann architecture~\cite{W22_qvn,W24_qvn} 
which treat Choi states as quantum stored-program states,
and use operations on Choi states for computation 
(see Appendix~\ref{sec:qvn} for more information).

% The above definition~(\ref{eq:codeerror}) does not specify what it given:
% $\Phi$ or its Choi state $\omega_\Phi$.
% They make a difference.

% We want to use the Choi form. 
% But when using Choi form, as Fig. 1 shows, 
% the tail part is assumed to be noise-free.
% This is not practical.
% Also, if in practice the Choi form is given, i.e., 
% an ebit is given and only one part of it is noisy,
% we can simply do stabilizer measurement and correction, 
% which is a Bell measurement.
% It is global in this setting and can correct any noise without coding. 
% So, in order for the Choi form to be practical,
% we need to extend this as follows. 

\begin{figure}[t!]
    \centering
    \includegraphics[width=0.4\textwidth]{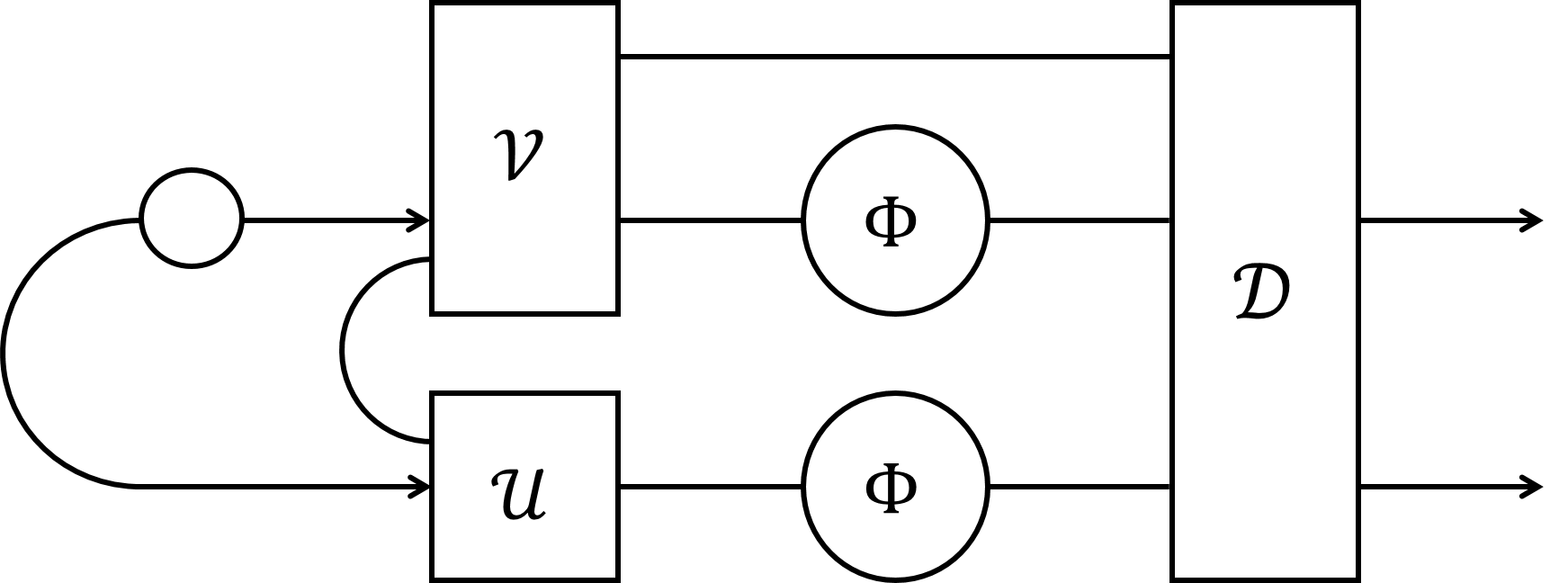}
    \caption{A schematic diagram of quantum Choi codings for a noise channel $\Phi$:
    the encoding is a superchannel $\hat{\C S}_E$ which is a bipartite operation involving 
    a `pre-' operator $\C U$ and a `post' operator $\C V$, 
    and the decoding is a channel $\C D$.
    The input source states are Choi states (on the left with a small circle), 
    and for the quantum control-assisted protocols
    there is also a noiseless register (the upper wire).}
    \label{fig:choicode}
\end{figure}

Therefore, we introduce coding models for Choi states (see Fig.~\ref{fig:choicode}) and obtain 
the following two theorems as our central results. 
The encoding is restricted to be superchannels in order to preserve the Choi form,
but the decoding can be general channels. 
%This will use entangling gates acting on the two parts of Choi states.
%Such gates can be simulated by quantum gate teleportation
%consuming ebits and LOCC operations. 
%We do not consider classical communication between 
%the encoding and decoding. 
Define the Choi coherent information for a channel $\Phi$ with input state $\omega_\C E$ as 
$I_c(\omega_\C E,\Phi^{\otimes2}):=S(\Phi^{\otimes2}(\omega_\C E))-S((\Phi^{c})^{\otimes2}(\omega_\C E))$,
%with $\Psi$ as the complementary channel of $\Phi$. 
the Choi mutual information as 
$I(\omega_\C E,\Phi^{\otimes2})=S(\omega_\C E)+I_c(\omega_\C E,\Phi^{\otimes2})$.

\begin{theorem}
    (Choi capacity of a quantum channel) 
For a quantum channel $\Phi$, a Choi coding protocol contains an encoding superchannel $ \hat{\C S}_E$
and a decoding channel $\C D$ so that 
\be F(\omega^{\otimes 2k},(\C D \Phi^{\otimes 2n} \hat{\C S}_E \otimes \I^{\otimes 2k})(\omega^{\otimes 2k}))
\geq 1-\epsilon, \label{eq:fchoicapacity}\ee 
for positive integers $n$ and $k$ and any $\epsilon\in [0,1]$,
and the Choi capacity, which is the supremum of all achievable rate $\alpha$
for $k=\lfloor \alpha n \rfloor$, is 
    \be \hat{Q}(\Phi)=\lim_{n\ra \infty }\max_{\omega_\C E} \frac{1}{2n}I_c(\omega_\C E,\Phi^{\otimes 2n}). \ee 
\end{theorem}

The above condition~(\ref{eq:fchoicapacity}) for the coding fidelity means $\C D \Phi^{\otimes 2n} \hat{\C S}_E $ 
is close to an identity channel $\I$.
For any input Choi state, the encoded state is also a Choi state,
so that the input to the second half of the noise channel $\Phi^{\otimes n}$ is a completely mixed state.
This is different from the usual quantum coding protocols. 
For the decoder channel $\C D$, it can also be 
simulated by a superchannel together with classical communication among 
the two ports of the output Choi states.

% the fidelity between 
% $\omega_\C E$ and $\C D \Phi^{\otimes 2n} \hat{\C S}_E (\omega_\C E)$ is close to 1. 
% %\be F(\omega_\C E,\C D \Phi^{\otimes 2n} \hat{\C S}_E (\omega_\C E)\geq 1-\epsilon, \label{eq:codeerror}\ee 

%since we can choose $\omega_{\Phi}$ as Bell states, 
%which form a basis for arbitrary state. 
%The fidelity function is jointly concave. 
% but this only hold for mixture of Bell states not superposition. 

% Even the input $\omega_{\Phi}$ is only finite but spanning, 
% the $\hat{\C S}_D \Phi^{\otimes 2n}$ is still close to $\I$.

The operational meaning of Choi mutual information is provided by the
quantum control-assisted (CA) Choi coding protocols,
which is the analog of usual EA protocols.
We name it as CA since it naturally corresponds to the 
quantum control unit in von Neumann architecture. 

%The quantum control units are nothing but the ancilla used for the dilation
%of a source program channel $\C E$.

% What is the operational meaning of Choi mutual information?
% We show that it is the 
% quantum control-assisted Choi channel capacity. 
% Using quantum control unit it is natrual the capacity gets larger. 
% We also call it as quantum vN channel capacity.
% It is a mutual information, so it is easy to show it is subadditive 
% and the capacity is single-letter form.

\begin{theorem}
    (Quantum control-assisted Choi capacity of a quantum channel) 
For a quantum channel $\Phi$, a quantum control-assisted Choi coding protocol 
contains an encoding superchannel $ \hat{\C S}_E$
and a decoding channel $\C D$ so that 
\be F(\omega^{\otimes 2k},\C D (\Phi\otimes \I)^{\otimes 2n} \hat{\C S}_E \otimes \I^{\otimes 2k})(\omega^{\otimes 2k}))
\geq 1-\epsilon, \ee 
for positive integers $n$ and $k$ and any $\epsilon\in [0,1]$,
and the CA Choi capacity, which is the supremum of all achievable rate $\alpha$
for $k=\lfloor \alpha n \rfloor$, is 
    \be \hat{Q}_{CA}(\Phi)=\max_{\omega_\C E} \frac{1}{4} I(\omega_\C E,\Phi^{\otimes2}).  \ee 
\end{theorem}

%Note there is a factor of $\frac{1}{2}$ compared with the usual capacity 
%since $2n$ copies of the channel $\Phi$ instead of $n$ are used. 

We now explain the proofs, with the details presented in the Appendix~\ref{sec:proof}. 
A proof includes a direct theorem and a converse theorem,
with the former shows a rate with the claimed expression for a capacity is achievable, 
while the latter shows that a capacity cannot exceed it.
The proof follows a systematic method that can be applied 
to all the four quantum capacities mentioned in this work.
Traditionally, 
the quantum and EA capacities are established using different methods.
Here, we find that the proof for the usual quantum capacity $Q$ can be 
extended to the EA capacity $Q_{EA}$, hence 
can also be adopted to $\hat{Q}$ and $\hat{Q}_{CA}$.
This is based on the following 
observation of the quantum error correction (QEC) condition and Hamming bound. 

Given a code space projector $P$ and a set of Kraus operators $E_i$ representing 
a noise channel $\Phi$,
the exact QEC condition is 
\be PE_i^\dagger E_j P=c_{ij}P, \ee 
for (the transpose of) the matrix $[c_{ij}]$ as nothing but the environment state $\rho_E$~\cite{KL97}.
By diagonalizing it, this leads to $PF_i^\dagger F_j P=p_i\delta_{ij}P$, 
with effective errors $F_i$ whose action is actually unitary~\cite{NC00}. 
The rank of $\rho_E$ can be smaller than the cardinality of the index $i$,
leading to the notable degeneracy phenomena for quantum codes~\cite{Got98,SS07}.
So the number of correctable errors $N$,
as a measure of the noise of the channel, is determined by the entropy of $\rho_E$, $S_E$.
This is distinct from the classical case, 
for which the condition is $\langle a|E_i^\dagger E_j |b\ket=\delta_{ab}\delta_{ij}p_{ia}$
for $p_{ia}\geq 0$,
that is, it only holds for classical states, and the errors are orthogonal by themselves.
The noise of the channel is determined by the conditional entropy $S_{Y|X}$
between the input space $X$ and output $Y$.
For the EA capacity, the mutual information $I(X:Y)$ is actually also a coherent information 
by noting that the output entropy not only includes $S_Y$,
but also the assisted entanglement, which is about $\frac{1}{2}I(X:E)$~\cite{Bow02,BDH+14}, 
leading to the EA capacity as $\frac{1}{2}I(X:E)+S_Y-S_E=\frac{1}{2}I(X:Y)$,
cf. Eq.~(\ref{eq:capacityea}).

For a system of dimension $M$, logical dimension $K\leq M$, a number of correctable errors $N$,
the Hamming bound requires $KN\leq M$~\cite{AC97,Kle07}. %(cite Klesse, Cerf paper)
This is also the underlying idea for the packing lemma
which is used to prove the direct theorem~\cite{Wil17}. 
In more details, 
the proof for the direct theorem is to first show that 
a measure with the `flat' input (completely mixed state)
is a lower bound for a capacity~\cite{Kle07,Wat18}. 
This is to show the coding fidelity is close to 1 
by using unitarily invariant ensemble of codes. 
The accuracy parameter $\epsilon$ is upper bounded by $(\frac{KN}{M})^{1/2}$,  
%$\sqrt{\frac{KN}{M}}$ $\sqrt{KN/M}$,
whose values are summarized in the Table~\ref{tab:knm} for the four types of quantum capacities. 
In the large-$n$ setting, 
with the flat input being an i.i.d. source,
also i.i.d. for the receiver,
one only considers the correction of typical Kraus operators
and the typical subspace at the receiver. 
Then this extends to the case of flat input within any typical subspace, 
and then uses the well-known formula 
\be S(\rho)= \lim_{\varepsilon\ra 0} \lim_{n\ra \infty} 
\frac{1}{n} S(\pi_{\varepsilon,n})\ee 
for $\pi_{\varepsilon,n}$ as the flat state on the $\varepsilon$-typical subspace of $\rho$~\cite{BSST02}, this generalizes to any input states allowed in a coding protocol.

\begin{table}[t!]
    \centering
    \begin{tabular}{c|c|l} 
    \hline 
     Capacity    & Channel & Parameters   \\ \hline 
      $Q$   & $\Phi$ & $(K,N,M)$ \\ \hline 
      $Q_{EA}$   & $\Phi\otimes \I$ & $(K^2,N,M^2)$ \\ \hline 
      $\hat{Q}$   & $\Phi^{\otimes 2}$ & $(K^2,N^2,M^2)$ \\ \hline 
      $\hat{Q}_{CA}$   & $(\Phi\otimes \I)^{\otimes 2}$ & $(K^4,N^2,M^4)$ \\ \hline 
    \end{tabular}
    \caption{The channel form and parameters for the proof of the direct theorem 
    for each capacity, i.e., a measure with the flat input (completely mixed state)
as a lower bound. 
The actual logical space dimension for all cases is $K$, leading to a factor of $\frac{1}{2}$
for the expression of $Q_{EA}$ and $\hat{Q}$, and $\frac{1}{4}$ for $\hat{Q}_{CA}$.}
    \label{tab:knm}
\end{table}

% For the EA case, we have $\Phi$ replaced by $\Phi\otimes \I$,
% $K$ replaced by $K^2$, $M$ replaced by $M^2$.
% The source entropy is also passed to the receiver,
% and the factor of $1/2$ is due to the fact that
% the protected information is a subspace of the total input source. 

% For classical codes, $N$ relates to the size of space of conditional typical strings,
% For quantum codes, $N$ relates to typical Kraus operators for 
% the large-$n$ setting $\Phi^{\otimes n}$.

The proof for the converse theorem is to relate each coding model to an entanglement generation task,  
and show that the capacity of this task is upper bounded 
by the claimed measure 
relying on the continuity of entropy and the data-process inequality~\cite{NC00}.

We can also obtain classical analog of the above two capacities,
by using mixture of Choi states as the source,
and use Holevo's approach to establish classical capacities~\cite{Wat18}.
Let 
$\omega_\C E'=\sum_i p_i |i\ket\bra i| \otimes \Phi^{\otimes 2n}(\omega_{\C E_i})$
and $\omega_\C E=\sum_i p_i \omega_{\C E_i}$,
the classical Choi capacity is 
\be \hat{C}(\Phi)= \lim_{n\ra \infty }\max_{\{p_i, \C E_i\}} \frac{1}{2n}\chi(\omega_\C E,\Phi^{\otimes 2n}), \ee 
for $\chi$ as the Holevo mutual information of $\omega_\C E'$,
and the quantum control-assisted classical Choi capacity is
\be  \hat{C}_{CA}(\Phi)= \max_{\omega_\C E} \frac{1}{2}I(\omega_\C E,\Phi^{\otimes 2}). \ee 
The expected relation $\hat{C}_{CA}(\Phi)=2\hat{Q}_{CA}(\Phi)$
can be proven by quantum teleportation and superdense coding (see Appendix~\ref{sec:qcchoi}).
From their formulas, it is clear that 
\be Q\geq  \hat{Q}, \; Q_{EA}\geq  \hat{Q}_{CA},\ee 
\be C\geq  \hat{C}, \; C_{EA}\geq \hat{C}_{CA},\ee 
as the Choi coding protocols are more restrictive.

Furthermore, the channel-state duality can be used iteratively 
leading to higher-order quantum operations~\cite{CDP08}.
This generates a chain of capacities for each of the four types of capacities above.
For instance, let channels be the 1st order states and for the $t$-th order states $\omega_{\C E^{(t)}}$,
denote $I_{c_r}^{(t)}(\Phi):=\lim_{n\ra \infty}\max_{\C E^{(t)}}\frac{1}{n2^t}I_c(\omega_{\C E^{(t)}},\Phi^{\otimes n 2^t})$, $I_{c_r}(\Phi):=\max_\rho I_{c_r}(\rho,\Phi)$,
then it holds 
\be I_c(\pi,\Phi)\leq \cdots \leq I_{c_r}^{(t)}(\Phi) \leq \cdots I_{c_r}^{(1)}(\Phi) \leq I_{c_r}(\Phi),\ee
and the lower bound $I_c(\pi,\Phi)$ serves as the quantum capacity 
when the encoding is unital~\cite{WLWL24},
for $\pi$ denoting the flat input state.
For degradable channels~\cite{DS05} the regularization is unnecessary and 
the above chain reduces to 
\be I_c(\pi,\Phi)\leq \cdots \leq I_{c}^{(t)}(\Phi) \leq \cdots I_{c}^{(1)}(\Phi) \leq I_{c}(\Phi),\ee
for $I_{c}^{(t)}(\Phi):=\max_{\C E^{(t)}} \frac{1}{2^t} I_{c}(\omega_{\C E^{(t)}},\Phi^{\otimes 2^t})$.
It remains to explore further the gaps in the chain, especially the gap
between $I_{c}(\Phi)$ and $I_{c}^{(1)}(\Phi)$,
and under what conditions the gaps can collapse.

This enriched family of capacities 
is consistent with an early unifying approach to characterize quantum capacities~\cite{HHH00}
and the recent resource-theoretic approach~\cite{WLWL24},
by highlighting the roles of information source and encoding,
and uncovers the fundamental role of coherent information.
It could help for the understanding of the novel
phenomena of superadditivity~\cite{SS07,SY08,LLS18,LLS+23},
with examples studied in Appendix~\ref{sec:example}. 

%of coherent information and Holevo quantity
% The Choi coherent information shows different behavior from coherent information,
% similarly for the classical measures, 
% and examples showing superadditivity~\cite{SS07,SY08,LLS18,LLS+23} 
% does not necessarily work for Choi-type measures
% as the input source states are required to be Choi states~\cite{LW25}.   

% We were unable to show Choi coherent information is subadditive. 
% However, using it we obtain interesting results for channels 
% that show superadditivity, such as super-activation, super-amplification.
% show in appendix. 
% This is a kind of ``renormalization''. 

%The control-assistance is a special type of entanglement-assistance.

% it can be viewed as a particular quantum coding with Choi states as input to the channel,
% and the encoded states are still Choi states. 

% and it can also be viewed as a two-stage quantum comb 
% in order to protect the source Choi state $\omega_{\Phi}$ 
% (by bending over the ebits and identify tail (head) as input (output) via ISI scheme).

\section{Applications}

% \begin{figure}[h!]
%     \centering
%     \includegraphics[width=0.4\textwidth]{sc}
%     \caption{superchannel (right) and ISI scheme (left)}
%     \label{fig:sc}
% \end{figure}

% explain application of Choi/CA codings, justify the noise-free condition for CA.  
% The noise-free ancilla comes from a pre-step that establish a remote purified Choi state 
% between A and B. A holds the ancilla a while B holds the Choi state. 
% Then A do a 2nd-step encoding by coupling the ancilla a with his program and do the encoding,
% so the post unitary in the super-isometry is actually two-step (similar with EA):
% then the noise-free condition is reasonable as it is held by Bob. 
% But we can simply treat the post unitary as a single step.
% another setting for CA is to use LCU of post unitary, so the control only exists for a while,
% and there is no need to do QEC on it. 

For the applications of the Choi coding models we developed above, 
the quantum von Neumann architecture is a suitable setting~\cite{W22_qvn}. 
The Choi coding can be used to protect quantum program states in the memory unit
which are Choi states. 
The primary resources are ebits whose quality can be 
guaranteed by entanglement purification or error-correction schemes
involving stabilizer measurements~\cite{NC00}. 
Given a program state $\omega_{\Phi}$, 
the action of the channel $\Phi(\rho)$ can be obtained 
via the initial-state injection (ISI) scheme (see Appendix~\ref{sec:qvn}).  
Namely, by constructing a measurement from $\rho$, 
the expectation value $\text{tr}(\C A \Phi(\rho))$ for any observable $\C A$ 
can be read out from $\omega_{\Phi}$~\cite{W20_choi}.

% After a superchannel, we have 
% \be \hat{\C S} (\Phi)(\rho)=\text{tr}_{\bar{\text{A}}} \C V \otimes \tilde{\C U} 
% (\omega_{\Phi} \otimes \omega) (\I\otimes \rho^t\otimes |0\ket \bra 0|), \ee
% where the support of each operator shall be easy to see hence omitted for simplicity.
% The trace is over the subsystems except the top one, A.
% The unitary $\tilde{\C U}$ is the transpose of $U$ conjugated by a swap. 

% Given a program $\omega_{\Phi}$, an initial input state $\rho$ is written into the tail 
% by a binary measurement 
% \be \{M_0=\sqrt{\rho^t}, M_1=\sqrt{\I-\rho^t}\}, \ee 
% and the final output in terms of an 
% expectation value $\text{tr}(\C A \Phi(\rho))$ for observable $\C A$ 
% can be read out from the head~\cite{W20_choi}.
% Note the superscript $t$ is the transpose. 
% This leads to the universality to simulate any quantum algorithm 
% in the usual circuit model whose input needs to be prepared first~\cite{W22_qvn}. 
% For the outcome 0, the correct answer is obtained. 
% For the outcome 1, the offset $\text{tr}(\C A \Phi(\I))/d$ can be deleted to get the correct result.
% This is referred to as the initial-state injection (ISI) scheme. 

The CA Choi coding assumes a noise-free control register,
which, similar with the EA setting, 
could come from a pre-round of error correction.
This coding is proper to enable the memory and control units jointly fault-tolerant.
As explained in the Appendix~\ref{sec:qvn}, 
the quantum control unit is formed by on-demand qubits 
whose coherence time can be short.
An explicit CA Choi coding scheme is to use the form of 
the linear combination of unitary operators
$\sum_ic_i U_i$~\cite{BCC+15} 
for the post operator in the super-isometry encoding (see Fig.~\ref{fig:choicode}),
with the amplitudes $c_i$ stored in the control unit,
and the gates $U_i$ stored in the memory unit. 
The decoder involves a projection on the control qubits,
whose success probability can be boosted by using 
the oblivious amplitude amplification algorithm~\cite{BCC+14}.

%explain LCU, OAA and ISI 

%\section{code design}

To conclude, in this work we introduced new types of channel capacities 
motivated by the channel-state duality. 
These capacities enrich the content of quantum Shannon theory,
and could benefit the study of superadditivity,
the design of fault-tolerant schemes and error-correction codes. 

% Finally, we observe that the Choi coding models 
% inspire new types of stabilizer codes~\cite{Got98} that differ from 
% traditional codes or EA codes~\cite{LB13}.
% The stabilizers come from $Z$-type stabilizers 
% and stabilizers of ebits, 
% which are pairs of $XX$ and $ZZ$,
% mapped by the encoding operations. 
% It remains to see how these different types of stabilizer codes compare in various tasks. 

% It is also important to design QEC codes following Choi coding models.
% Here we describe a stabilizer code which is an extension of the four-qubit code,
% which is the smallest code for qubit erasure channels.
% Our code is a four-qudit code which encodes a unitary gate $U=[u_{ij}]$, or 
% via its Choi state $|U\ket=\sum_i |u_i\ket |i\ket$, as 
% \be |U\ket \mapsto \sum_{ijl} u_{ij} |j\ket |j+l\ket |i+l\ket |i\ket \ee 
% using qudit entangling gate and a qudit ebit as ancilla.
% If any of the four qudit is lost,
% it is not hard to check that $|U\ket$ can be recovered from the rest. 
% A state $|u_i\ket$ can be obtained by making a projection $|i\ket\bra i|$ on it. 
% The qudit can be viewed as a collection of qubits, 
% therefore, this code is a nontrivial code for qubit erasure channels.
% In general, a Choi-type stabilizer code is generated by 
% the stabilizers of ebits (which are pairs of $XX$ and $ZZ$)
% mapped by the encoding operations. 
% The decoding has to be global, with respect to the bipartition of Choi states,
% as the stabilizer measurements 

%stabilizer codes, 
%A class of erasure codes, 

\section{Acknowledgement}
This work has been funded by
the National Natural Science Foundation of China under Grants
12047503 and 12105343.
Suggestions from M. Hayashi, K. Li, S. Luo, 
M. Wilde, and Y.-J. Wang are acknowledged.

\appendix 

% \begin{paracol}{2}
% 	\switchcolumn
% 	%\begin{leftcolumn}
% 		gfdsvsdv
% 	%\end{leftcolumn}
% 	%\switchcolumn
% 	%\begin{rightcolumn}
% 		dgafsva
% 		\lipsum[1-1]
% 	%\end{rightcolumn}
% 	\switchcolumn
% \end{paracol}

%\newpage

\section{Representations of quantum superchannels}
\label{sec:superc}

Here we describe the representations of quantum superchannels in more details. 
Recall that the channel-state duality~\cite{Cho75,Jam72} states that a channel 
$\Phi: \C D(\C H_1) \ra \C D(\C H_2)$ 
can be represented as a Choi state 
\be \omega_{\Phi} := \Phi \otimes \I (\omega), \ee 
for $\omega:=|\omega\ket\bra \omega|$ as the Bell state.
It is clear to see the partial trace over the input and output space  
yields $\text{tr}_1 \omega_{\Phi}=\Phi(\I) /d$,
and $\text{tr}_2 \omega_{\Phi}= \I /d$,
for $d$ as the input space dimension. 
Therefore, Choi states form a convex subset 
$\C C(\C H_1\otimes \C H_2) \subset \C D(\C H_1\otimes \C H_2)$
of the set of bipartite states. 
The Kraus operators can be found from the eigenvalue decomposition of $\omega_{\Phi}$,
and the rank of the channel is the rank of $\omega_{\Phi}$.
From dilation, the channel $\Phi$ can be represented as a unitary $U$
requiring an ancilla, and the resulting tripartite state 
is a `purified Choi state', $|\phi_\Phi\ket$,
which is a purification of the Choi state $\omega_{\Phi}$.

A superchannel $\hat{\C S}: \C C(\C H_1 \otimes \C H_2) \ra \C C(\C H_{1'} \otimes \C H_{2'})$ 
that acts on a channel $\Phi$ can be realized as
\be \hat{\C S} (\Phi)(\rho)= \text{tr}_{a_1a_2} \C U_2 \; (\Phi \otimes \I)\; 
\C U_1 (\rho\otimes |00\ket \bra 00|),  \ee
for $\rho \in \C D(\C H_{1'})$, 
$\C U_1$ and $\C U_2$ are unitary operators,
and $a_1$ is an ancilla required by $\C U_1$,
$a_2$ is an ancilla further required by $\C U_2$,
and their initial state is $|00\ket$.
Note this is equivalent to Eq.~(\ref{eq:superchannel}) in the main text,
and we use $\C U$ to denote the superoperator form of $U$.
The dimension of $\C U_2$ could be larger than that of $\C U_1$
due to the additional ancilla $a_2$~\cite{WW23}.
%but we can enlarge the later so that they 
It can also be written as the Kraus operator-sum form 
\be \hat{\C S}(\omega_{\Phi})=\sum_\mu S_\mu \omega_{\Phi} S_\mu^\dagger\ee
with Kraus operators
\be S_\mu= \sum_m K_{2,m\mu} \otimes K_{1,m} \label{eq:s}\ee 
and $\sum_\mu S_\mu^\dagger S_\mu=\I$, 
$K_{1,m}$ and $K_{2,m\mu}$ are Kraus operators from the pre- and post unitary operators with
$K_{1,m}=\bra m|U_1|0\ket$ and $K_{2,m\mu}=\bra m|U_2|\mu\ket$. 
It is also easy to see 
the new channel $\hat{\C S} (\Phi)$ is represented by Kraus operators
\be F_{i\mu}= \sum_m K_{2,m\mu}K_i K_{1,m}^t,\ee
with $\sum_{i\mu} F_{i\mu}^\dagger F_{i\mu}=\I.$
The superscript $t$ is matrix transposition.
%The superscript $\mathsf{t}$ is matrix transposition.
The sum over $m$ signifies the quantum memory between $U_1$ and $U_2$,
which is the state from the ancilla $a_1$.

From the channel-state duality, we know $\omega_{\Phi}$ is obtained
by `bending over' the input space of $\Phi$.
When $\omega_{\Phi}$ is given, the realization of $\hat{\C S}(\omega_{\Phi})$
requires a projection $|0\ket \bra 0|$ on the ancilla $a_1$.
Simply tracing out it also leads to a superchannel which is a mixture of 
$\hat{\C S}_i$ for each projection $|i\ket \bra i|$.  
Therefore, when using Choi states as the input source
one can ignore the final projection on the ancilla 
for the realization of a superchannel. 
In total, a superchannel can also be viewed as a `super-isometry' 
$V:\C H_1\otimes \C H_2 \ra \C H_1\otimes \C H_2 \otimes \C H_{a_1}\otimes \C H_{a_1}\otimes \C H_{a_2} $
for the initial states of (two copies of) ancilla $a_1$ as ebits 
and ancilla $a_2$ as $|0\ket$,
together with the final tracing on them. 
We can also ignore $a_2$ for simplicity or by extending the pre- unitary $U_1$
to be of the same dimension with $U_2$. 

\begin{figure}[t!]
    \centering
    \includegraphics[width=0.3\textwidth]{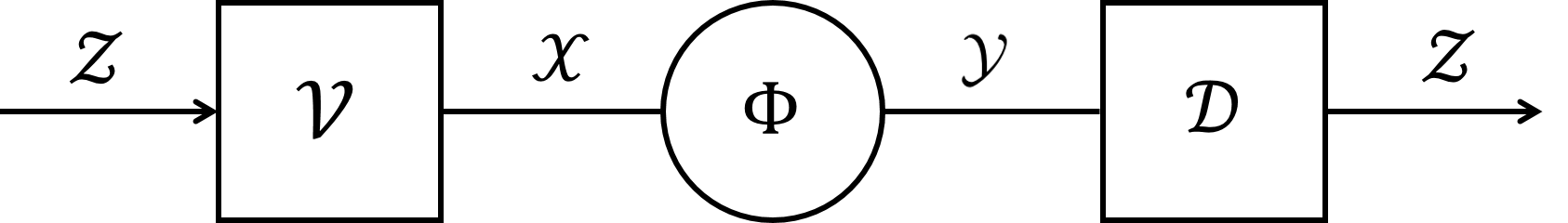}\vspace{0.5cm}
    \includegraphics[width=0.3\textwidth]{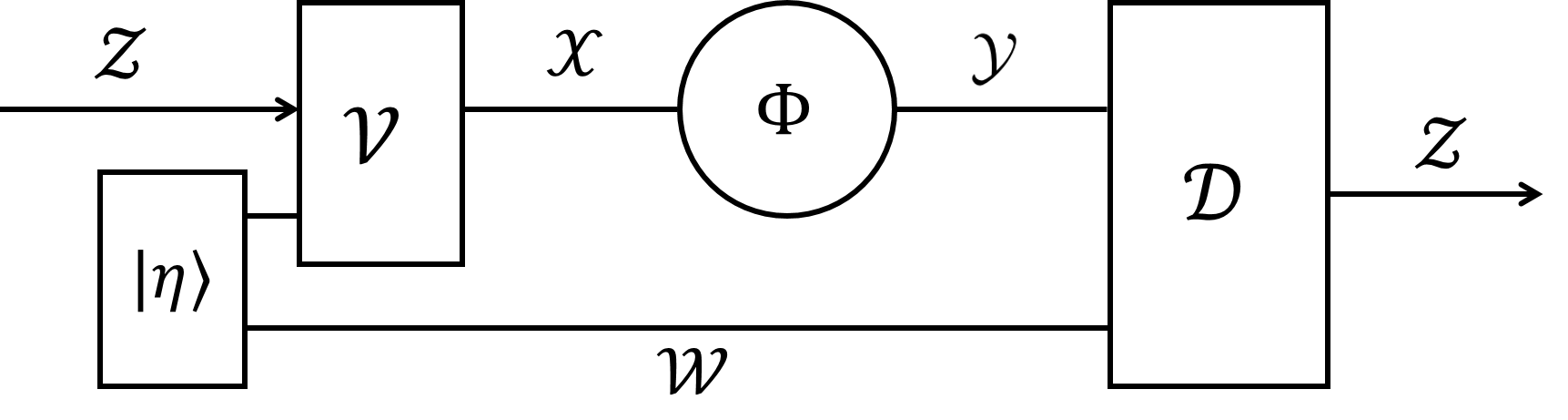}
    \caption{A schematic diagram of the standard quantum coding (top) and EA coding (bottom).}
    \label{fig:code12}
\end{figure}

With respect to the bipartition of a Choi state,
a superchannel is `semi-local' due to the usage of ebits. 
In the setting of error correction, a local scheme without ebits can describe
the usual setting of error correction without assistance,
with a pre- encoding isometry $V$
and a post decoding channel $\C D$.
While for the EA case, the encoding is two-stage $(V\otimes \I)|\eta\ket$ 
with $|\eta\ket$ as an assisted entangled state and an isometry $V$ 
that does not act on the noise-free quantum memory (see Fig.~\ref{fig:code12}). 
Note the trivial case shall be avoided,
which is to simply swap the input source with the noise-free ancilla in the superchannel,
and the EA setting indeed avoids this.
Furthermore, if forward and backward classical communications between the two parts
are allowed, together with ebits this can be used to realize any global operations on Choi states,
which may not preserve the form of Choi states. 
For instance, a CNOT gate can convert an ebit to a product state.
However, in the definition for the Choi coding models,
we allow any operations for the decoder as long as they preserve the Choi form.

\section{Quantum von Neumann architecture}
\label{sec:qvn}

Here we describe some primary operations in the recently introduced 
quantum von Neumann architecture~\cite{W22_qvn,LWLW23}. 
Compared with the usual quantum circuit model,
it explicitly includes quantum memory unit as Choi states 
and quantum control unit as qubits. 

In this setting, a Choi state is also called a quantum program. 
Given a program $\omega_{\Phi}$, a basic operation is to recover its action on state $\Phi(\rho)$.
This can be achieved as 
\be \Phi(\rho)= d \; \text{tr}_1 [\omega_{\Phi} (\I \otimes \rho^t) ], \label{eq:readout}\ee
for $\rho^t$ as the transpose of a state $\rho \in \C D(\C H_1)$
and the trace is on the input space $\C H_1$.
This can be realized as 
% After a superchannel, we have 
% \be \hat{\C S} (\Phi)(\rho)=\text{tr}_{\bar{\text{A}}} \C V \otimes \tilde{\C U} 
% (\omega_{\Phi} \otimes \omega) (\I\otimes \rho^t\otimes |0\ket \bra 0|), \ee
% where the support of each operator shall be easy to see hence omitted for simplicity.
% The trace is over the subsystems except the top one, A.
% The unitary $\tilde{\C U}$ is the transpose of $U$ conjugated by a swap. 
a binary measurement 
\be \{M_0=\sqrt{\rho^t}, M_1=\sqrt{\I-\rho^t}\}, \ee 
and the final output in terms of an 
expectation value $\text{tr}(\C A \Phi(\rho))$ for observable $\C A$ 
can be read out from the output space~\cite{W20_choi,W22_qvn}.
%Note the superscript $t$ is the transpose. 
%This leads to the universality to simulate any quantum algorithm 
%in the usual circuit model whose input needs to be prepared first~\cite{W22_qvn}. 
For the outcome 0, the correct answer is obtained. 
For the outcome 1, the offset $\text{tr}(\C A \Phi(\I))/d$ can be deleted to get the correct result.
This is referred to as the initial-state injection (ISI) scheme. 

Two programs can be fused together. 
This is based on quantum teleportation~\cite{NC00}  and its extensions.
To teleport an unknown state $|\psi\ket_S$ from system $S$ to register $B$,
it needs a Bell state $|\omega\ket_{AB}$ and Bell measurement $M_{AS}$ which yields 
Pauli byproduct $\sigma_{i}$ so that 
\be |\psi\ket_{B}= \sigma_{i,B}M_{AS}(i)
|\omega\ket_{AB}|\psi\ket_S.  \ee 
This can be extended to quantum gate teleportation~\cite{GC99} by modifying the byproduct
or the universal quantum teleportation (UQT) scheme~\cite{W20_choi} 
by modifying the measurement while keeping the Pauli byproduct. 
The UQT is based on the following symmetry 
\be U \sigma_i U^\dagger =\sum_j T_{ij} \sigma_j, \ee 
for general qudit case with $U\in$ SU($d$), 
and qudit Pauli unitary operators $\sigma_i$,
$T=[T_{ij}]\in$ SU($d^2$) is an affine representation of $U$~\cite{BZ06}.
By attaching $T$ to the Bell measurement circuit, the rotated measurement $M^T_\text{AS}$
enables  
\be U|\psi\ket_B= \sigma_{i,B}M^T_{AS}(i)
|\omega_{U^t}\ket_{AB}|\psi\ket_S,  \ee 
that is, the program $U$ is teleported or downloaded from $S$ to $B$. 
The transpose can be avoided by using 
$(U\otimes \I) |\omega\ket =(\I \otimes U^t) |\omega\ket$.
For unitary programs $|U_1\ket$ and $|U_2\ket$, 
they can be fused to yield $|U_1U_2\ket$ or $|U_2U_1\ket$.
For non-unitary programs $\omega_{\Phi_1}$ and $\omega_{\Phi_2}$,
it turns out a straightforward scheme is to use their purified Choi states $|\phi_{\Phi_1}\ket$ and $|\phi_{\Phi_2}\ket$, 
and attach ancillary ebits to store the ancillary states of them,
and use UQT to fuse them.

% Quantum teleportation has an important symmetry,
% This is manifested in the fact that the probability of its Pauli correction occurring is the same.
% The $X$ and $Z$ effects of Pauli on the input S,
% It can be expressed as Pauli's action on the final state.
% Its symmetry is $Z_d \times Z_d$,
% This is also the global symmetry of a one-dimensional graph\cite{W19_rev},
% We know that its application in computing is based on the quantum transmission mechanism.

% Arbitrary algorithmic process (e.g., $\bra \psi_f |). U_n \cdots U_2 U_1 |\psi_i \ket$)
% to prove the versatility of the model.
% In hardware, this mechanism can be used in the construction of quantum chips.

However, when a program is unknown, it cannot be downloaded exactly~\cite{NC97}.
We can use ISI scheme to obtain its action on particular observable. 
If approximation is allowed, 
then a more complicated scheme can be used~\cite{YRC20},
%It can be approximately downloaded by storing it in a more complicated form,
namely, using $U^{\otimes n}$ to act on a multipartite entangled state 
and a covariant measurement can download it approximately, 
with accuracy limited in the order $n^{-2}$.

Superchannels form another type of operations on programs 
which can convert one program into another.
This can be extended to a sequence of superchannels,
and together with quantum teleportation schemes, 
this leads to various operations and algorithms based on Choi states~\cite{LWLW23}. 

Another essential part of von Neumann architecture is the quantum control unit, 
which is formed by qubits that interact with data qubits.
They are ancillas but they carry out special functions,
therefore, it is necessary to treat them as a separate unit. 
Notable schemes include the quantum control over unknown gates~\cite{AFC14}, 
quantum switch~\cite{CAPV13}, 
the linear combination of unitary (LCU) operation algorithm~\cite{BCC+15} and 
oblivious amplitude amplification (OAA)~\cite{BCC+14}. 
Compared with the data unit, 
the control unit is of short circuit depth,
so different error-correction methods can be used.
%The CA Choi coding can be applied. 
%Note the decoding act both on the control and data units. 

Here we describe the LCU and OAA scheme that is developed in the setting 
of quantum simulation algorithm~\cite{BCC+14}. 
It is also a good example to describe the role of control unit. 
Using qubits can generate entanglement between the control and data units, 
and in particular, 
it can generate superposition of gates 
\be U=\sum_i c_i U_i,\ee 
however, this requires a post-selection on the control unit. 
To boost the success probability close to 1, 
the OAA scheme can be used which is an extension of amplitude amplification.
Namely, denote the whole action of LCU circuit as $W$,
and $W|0\ket |\psi\ket = \sqrt{p}|0\ket U |\psi\ket + \sqrt{1-p}|1\ket |\phi\ket$.
When a failure occurs, i.e. outcome 1,
the OAA will apply $S=WRW^\dagger R$ for $R$ as a reflection operator acting on the control unit,
and by application of $S^n$ for many iterations $n\in O(1/\sqrt{p})$,
the success rate can be boosted arbitrarily close to 1.

This will increase the circuit depth of the control unit, however.
A method that can maintain a constant circuit depth is to use OAA incoherently, 
that is, upon a failure one applies $S$ and performs the measurement,
and then repeat, and this scheme on average requires $O(1/p)$ iterations 
to boost the success rate close to 1. 

In the light of fault-tolerance, 
different error-processing schemes 
are suitable for the different units in a quantum von Neumann architecture. 
The Choi coding protocol is suitable for the quantum memory unit. 
The EA coding protocol is suitable for quantum communication.
For the quantum control unit,
a central feature or merit of it is that it is on-demand,
that is, the control qubits will be measured and refreshed quickly,
and they do not need to maintain a long coherence time. 
Instead of using powerful QEC codes, 
weaker schemes can be used, such as error suppression or approximate 
error correction schemes, 
including dynamical decoupling, approximate QEC codes,
or error-detection codes~\cite{LB13}. 
The basic idea is that when the control unit is needed, 
error detection or suppression on it is used, 
and when the short-depth circuit is finished, there is a high chance that 
errors do not occur or only accumulate slightly. 
The control qubits will be measured and refreshed again. 
The CA Choi coding protocol we developed in this work can be applied 
to jointly protect the quantum control and memory units.

\begin{figure}[t!]
    \centering
    \includegraphics[width=0.3\textwidth]{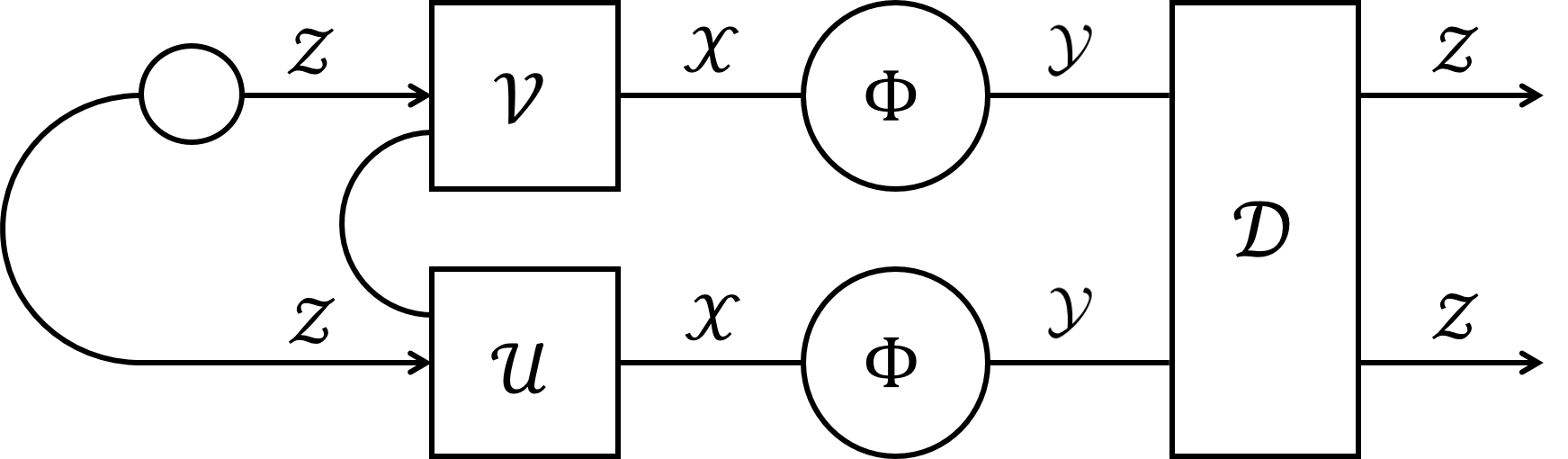}\vspace{0.5cm}
    \includegraphics[width=0.3\textwidth]{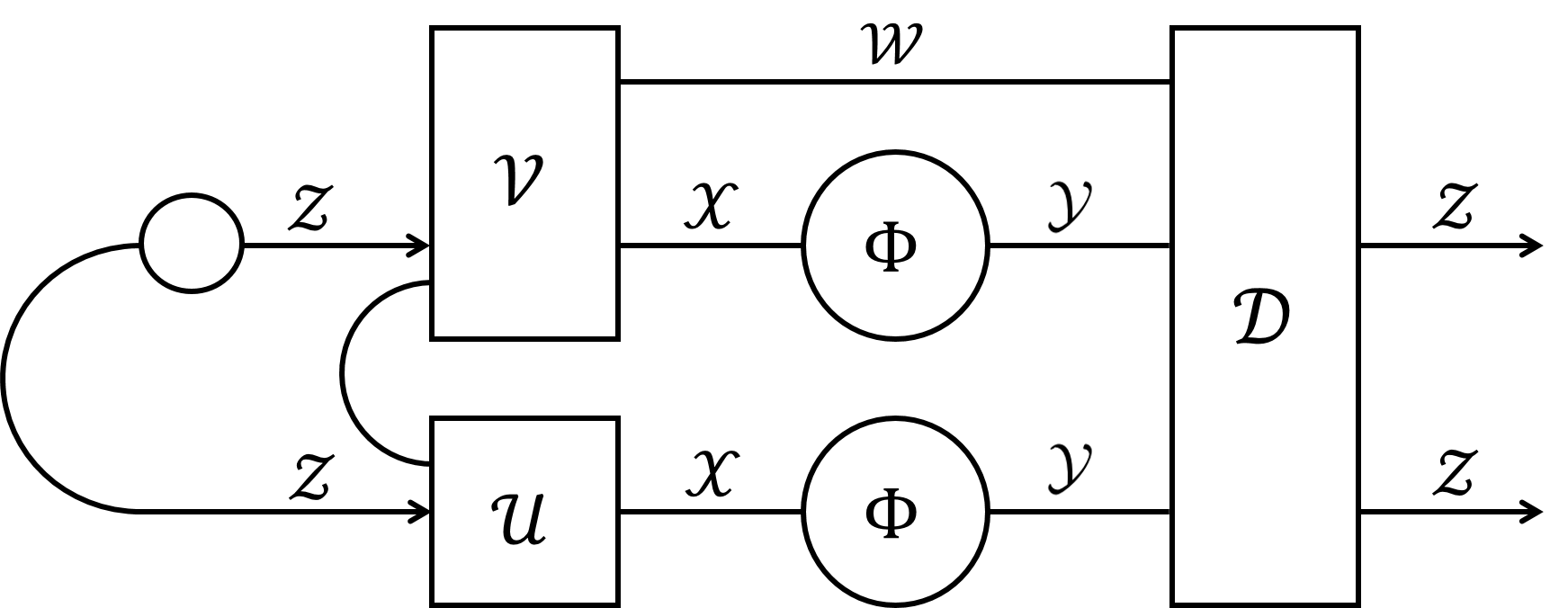}
    \caption{A schematic diagram of the quantum Choi coding (top) and CA Choi coding (bottom).}
    \label{fig:code34}
\end{figure}

\section{Proofs of the quantum channel capacities}
\label{sec:proof}

In this section, we present the details for the proofs of quantum channel capacities, 
which apply to the four types of quantum capacities in this work. 
We use $\C X, \C Y, \C Z, \dots$ to denote Hilbert spaces, 
and use $C(\C X, \C Y)$ to denote the set of channels $\Phi: \C D(\C X) \ra \C D(\C Y)$
acting on density matrices~\cite{Wat18}. 

First, we can define a quantum capacity in general as follows.
The difference for the four coding models 
is the coding protocol $\hat{\C S}$. 

\begin{definition}
(Quantum capacity of a channel) 
Let $\Phi\in C(\C X, \C Y)$ be a channel, and an integer $k=\lfloor \alpha n \rfloor$ 
for all but finitely many positive integers $n$ and an achievable rate $\alpha \geq 0$, 
there exists a superchannel $\hat{\C S}$
%channels $\Phi\in C(\C Z^{\otimes m}, \C X^{\otimes n})$ 
%and $\C D\in C(\C Y^{\otimes n}, \C Z^{\otimes m})$ 
such that 
\be F(\omega^{\otimes k},(\hat{\C S} (\Phi^{\otimes n})\otimes \I^{\otimes k})(\omega^{\otimes k}))
\geq 1-\epsilon, \label{eq:codeerror0}\ee 
%\be D_\diamond (\I^{\otimes m}, \C D \Phi^{\otimes n} \Phi) \leq \epsilon\ee
for every choice of a positive real number $\epsilon\in[0,1]$
and the quantum capacity of $\Phi$ is defined as the supremum of all $\alpha$.
\end{definition}

For the details of the original proofs for 
the quantum channel capacities, we refer to the books~\cite{Wil17,Wat18}.
% For the quantum channel capacity without assistance, 
% the direct theorem was due to ..,
% the converse theorem was due to ...
% For the quantum channel capacity with entanglement assistance, 
% the direct theorem was due to ..,
% the converse theorem was due to ...
For the direct theorem, the packing lemma plays the central role,
while the covering lemma is also involved in a proof 
for the quantum channel capacity without assistance~\cite{Dev05,Wil17}.
We have explained in the main text the QEC condition already requires 
the decoupling of environment states from the code, 
and using quantum packing lemma can deduce coherent information 
as the measure of quantum channel capacity. 
For the EA setting, it turns out it can be treated as a particular case of the above. 

Now we lay out a general framework for the proofs of the quantum channel capacities
following the logic of Refs.~\cite{Kle07,Wat18}.
We employ the random coding method for the direct theorem,
and the entanglement distribution scheme for the converse theorem. 
For the channel $\Phi$ with a set of Kraus operators $\{A_i\}$ and rank $N$,
denote the input/output system as $X/Y$ with dimension $M$, 
denote the purification system of the input as $R$ with dimension $M$,
the environment as $E$ with dimension $N$.
The input states are from a code subspace with projector $P$ and dimension $K$.
Denote a measure with input state $\rho$ to channel $\Phi$ as 
$m(\rho,\Phi)$, which would be a coherent information or mutual information. 
Also denote $\pi:=\I/M$ as the flat input state 
(i.e. a completely mixed state).

For the direct theorem, the first step is to 
show any rate $\alpha \leq m(\pi,\Phi)$ is achievable.
That is, for any such rate there exists a coding scheme 
such that the coding fidelity is arbitrarily close to 1. 
Using random coding scheme and the typicality for environment state $\rho_E$ and 
receiver state $\rho_Y$
can compute the lower bound of the fidelity.
Then this is generalized to $Q(\Phi)\geq m(\pi_C,\Phi)$ for $\pi_C$ as the flat state in 
any subspace of the input, and the BSST lemma
\be S(\rho)= \lim_{\varepsilon\ra 0} \lim_{n\ra \infty} 
\frac{1}{n} S(\pi_{\varepsilon,n})\ee 
converts $\pi_{\varepsilon,n}$ into arbitrary input state $\rho$ in the large-$n$ limit,
for $\pi_{\varepsilon,n}$ as the flat state on 
the $\varepsilon$-typical subspace of $\rho$~\cite{BSST02}.
Finally, it is straightforward to obtain $Q(\Phi)\geq m(\rho,\Phi)$.

We now present the theorem for the two Choi capacities 
$\hat{Q}$ and $\hat{Q}_{CA}$,
while the proof also applies to the other two quantum capacities
$Q$ and $Q_{EA}$.
For clarity, we define the two Choi capacities first. 

\begin{definition}
(Choi capacity of a quantum channel) 
Let $\Phi\in C(\C X, \C Y)$ be a channel, and an integer $k=\lfloor \alpha n \rfloor$ 
for all but finitely many positive integers $n$ and an achievable rate $\alpha \geq 0$, 
there exists a super-isometry $\hat{\C S}_E\in C(\C Z^{\otimes 2k}, \C X^{\otimes 2n})$ 
as the encoding and a decoding channel $\C D \in C(\C Y^{\otimes 2n}, \C Z^{\otimes 2k})$
%channels $\Phi\in C(\C Z^{\otimes m}, \C X^{\otimes n})$ 
%and $\C D\in C(\C Y^{\otimes n}, \C Z^{\otimes m})$ 
such that 
\be F(\omega^{\otimes 2k},\C D \Phi^{\otimes 2n} \hat{\C S}_E \otimes \I^{\otimes 2k})(\omega^{\otimes 2k}))
\geq 1-\epsilon, \label{eq:codeerror2}\ee 
%\be D_\diamond (\I^{\otimes m}, \C D \Phi^{\otimes n} \Phi) \leq \epsilon\ee
for every choice of a positive real number $\epsilon\in[0,1]$
and the Choi capacity of $\Phi$, denoted as $\hat{Q}(\Phi)$, 
is defined as the supremum of all $\alpha$.
\end{definition}

\begin{definition}
(CA Choi capacity of a quantum channel) 
Let $\Phi\in C(\C X, \C Y)$ be a channel, and an integer $k=\lfloor \alpha n \rfloor$ 
for all but finitely many positive integers $n$ and an achievable rate $\alpha \geq 0$, 
there exists a super-isometry $\hat{\C S}_E\in C(\C Z^{\otimes 2k},\C W^{\otimes 2n}\otimes \C X^{\otimes 2n})$ 
as the encoding and a decoding channel $\C D \in C(\C W^{\otimes 2n}\otimes \C Y^{\otimes 2n}, \C Z^{\otimes 2k})$
%channels $\Phi\in C(\C Z^{\otimes m}, \C X^{\otimes n})$ 
%and $\C D\in C(\C Y^{\otimes n}, \C Z^{\otimes m})$ 
such that 
\be F(\omega^{\otimes 2k},\C D (\Phi\otimes \I)^{\otimes 2n} \hat{\C S}_E \otimes \I^{\otimes 2k})(\omega^{\otimes 2k}))
\geq 1-\epsilon, \label{eq:codeerrorca}\ee 
%\be D_\diamond (\I^{\otimes m}, \C D \Phi^{\otimes n} \Phi) \leq \epsilon\ee
for every choice of a positive real number $\epsilon\in[0,1]$
and the CA Choi capacity of $\Phi$, denoted as $\hat{Q}_{CA}(\Phi)$, 
is defined as the supremum of all $\alpha$.
\end{definition}

We now state the main theorem as follows which includes 
the two theorems in the main text.
%Denote $I_{c_r}(\omega_\C E,\Phi^{\otimes 2})
%:=\lim_{n\ra \infty} \frac{1}{2n} I_c(\omega_\C E,\Phi^{\otimes 2n})$.

\begin{theorem}
    (Choi capacities of a quantum channel) 
For a quantum channel $\Phi\in C(\C X, \C Y)$, 
the Choi capacity is 
    \be \hat{Q}(\Phi)=
    \lim_{n\ra \infty} \frac{1}{2n} \max_{\omega_\C E} I_c(\omega_\C E,\Phi^{\otimes 2n}), \ee 
and the CA Choi capacity is 
 \be \hat{Q}_{CA}(\Phi)=\max_{\omega_\C E} \frac{1}{4} I(\omega_\C E,\Phi^{\otimes2}).\ee
\end{theorem}
\begin{proof}
(direct) We first prove the direct theorem.
We start from the usual quantum channel capacity $Q(\Phi)$ to illustrate the method.
%The goal is to prove any rate $\alpha \leq Ir$ is achievable. 
The first step is to show any rate $\alpha \leq I_c(\pi,\Phi)$ is achievable. 
That is, for any such rate there exists a coding scheme 
such that the coding fidelity is arbitrarily close to 1. 

The decoupling approach~\cite{SW02,HHW+08,ADHW09,Kle07} 
shows that there exists a decoding channel $\C D$ such that 
\be F \geq 1- \|\sigma_{RE}-\sigma_R\otimes \sigma_E\|_\text{tr} \label{eq:decoupl} \ee 
for state $\sigma_{RE}=\frac{1}{K}\sum_{ij} PA_i^\dagger A_j P \otimes |j\ket \bra i|$,
$\sigma_R=\frac{1}{K}P$, $\sigma_E=\frac{1}{K}\sum_{ij} \text{tr}(PA_i^\dagger A_j P)|j\ket \bra i|$.

Then using random coding scheme, consider the unitarily invariant ensemble of codes $UPU^\dagger$
for all $U$ acting on the input system following the Haar measure on the unitary group.
If the ensemble average $[F]$ is close to 1,
then there must exist a code with a fidelity $F$ which is close to 1~\cite{Kle07}.
Converting the trace norm into Frobenius norm due to $\|\cdot\|_\text{tr}\leq \sqrt{KN} \|\cdot\|_F$,
then the ensemble average of the error function $[\|\sigma_{RE}-\sigma_R\otimes \sigma_E\|_F]$
is computed using Weyl's theory of group invariants~\cite{Weyl46} to obtain
\be [F] \geq 1- \sqrt{KN} \|\Phi(\pi)\|_F. \label{eq:fbound} \ee 

In the large-$n$ setting, the channel is of the form $\Phi^{\otimes n}$.
As the input is the completely mixed state, 
the environment state is also a product state $\sigma_E^{\otimes n}$.
By diagonalizing $\sigma_E=[p_i]$, which corresponds to a canonical set of Kraus operators $\{A_i\}$ 
with $p_i=\text{tr}(A_i^\dagger A_i)/M$,
this forms an i.i.d. source so we can define $\varepsilon$-typical sequence of Kraus operators 
with size $N_\varepsilon \leq 2^{n (S_E+ \varepsilon)}$,
and only consider the correction of typical Kraus operators. 

At the receiver side, the decoder only consider the typical subspace 
with respect to the state $\Phi(\pi)$.
This relates $\|\Phi(\pi)^{\otimes n}\|_F$ to $S(\Phi(\pi))$. 
Then it is not hard to obtain for any rate $\alpha \leq I_c(\pi,\Phi)=S(\Phi(\pi))-S_E$, 
there exists a coding scheme 
such that the coding fidelity is arbitrarily close to 1.
So $Q(\Phi)\geq I_c(\pi,\Phi)$.

This generalizes to $Q(\Phi)\geq I_c(\pi_C,\Phi)$ for $\pi_C$ as any subspace flat state, 
by using $Q(\Phi)\geq Q(V \Phi) \geq I_c(\pi_C,\Phi)$ for $V$ as the isometric encoding from the subspace $C$ 
to the whole space. 
Finally, using the BSST lemma to obtain $Q(\Phi)\geq I_{c_r}(\Phi)$. %for any $\rho$.

This also extends to the EA setting. 
In an EA protocol (see Fig.~\ref{fig:code12}), 
a bipartite entangled state $|\eta\ket$ is needed and one part of it 
is feed into an encoding step, and the other is noise-free and sent to the receiver directly.
We treat the preparation of $|\eta\ket$ as an encoding stage of a fiducial system $X'$
with the same logical dimension with the input $X$, so that the whole encoding is two-stage $V=V_2V_1$,
and $P=VV^\dagger$.
For convenience,
we can let the noise-free part is of size $n$ so that the effective channel is $\Phi\otimes \I$. 
That is, the parameters $(K,N,M)$ that determines the fidelity now becomes $(K^2,N,M^2)$,
but the actual logical dimension is still $K$ (see Table~\ref{tab:knm} in the main text).

Then, in Eq.~(\ref{eq:decoupl}) the projection $P$ is from the two-stage encoding, 
the Kraus operators are $A_i\otimes \I$, and $K \mapsto K^2$.
For the unitarily invariant ensemble, it is constructed from two independent usages 
for the two stages in the encoding. 
%The fidelity bound in Eq.~(\ref{eq:fbound}) now becomes
%\be [F] \geq 1- K\sqrt{N/M} \|\Phi(\I/M)\|_F. \label{eq:fboundea} \ee 
In the large-$n$ setting, it is easy to see the number of typical Kraus operators
is still bounded by $2^{n (S_E+ \varepsilon)}$,
while the size of the typical subspace at the receiver is now bounded by 
$2^{n (S(\pi)+S(\Phi(\pi))+ \varepsilon)}$.
This leads to that the achievable rate 
$\alpha \leq \frac{1}{2}I(\pi,\Phi)=\frac{1}{2}(S(\pi)+S(\Phi(\pi))-S_E)$.
Finally, it is straightforward to obtain $Q_{EA}(\Phi)\geq I(\rho,\Phi)/2$ for any $\rho$.

% This generalizes to $Q_{EA}(\Phi)\geq I(\pi_C)/2$ for $\pi_C$ as any subspace, 
% by using $Q_{EA}(\Phi)\geq Q_{EA}(V \Phi) \geq I(\pi_C)/2$ 
% for $V$ as the isometric encoding from the subspace $C$ 
% to the whole space. 

Now for the Choi coding case, 
it is almost two copies of the case for the usual quantum coding.
In Eq.~(\ref{eq:decoupl}) the states are $\sigma_{RE}^{\otimes 2}$ and 
$\sigma_R^{\otimes 2}\otimes \sigma_E^{\otimes 2}$.
For the random coding, we use two independent actions on the two ports of the Choi states.
The parameters that determine the coding fidelity is $(K^2,N^2,M^2)$.
Then following the similar process it finally yields  
$\hat{Q}(\Phi)\geq 
\lim_{n\ra \infty} \frac{1}{2n} \max_{\omega_\C E} I_c(\omega_\C E,\Phi^{\otimes 2n})$. 
%for any $\omega_\C E$.
%$\hat{Q}(\Phi)\geq I_{c_r}(\omega_\C E,\Phi^{\otimes 2})$ for any $\omega_\C E$.

For the CA Choi coding case, 
it is similar with the EA coding, but instead of using a two-stage encoding, 
the encoding here is similar with the Choi coding case which is 
two independent parallel encoding forming a super-isometry, 
while additional noise-free ancilla is allowed in the post unitary of the encoding. 
%In Eq.~(\ref{eq:decoupl}) the states are $\sigma_{RE}^{\otimes 2}$ and 
%$\sigma_R^{\otimes 2}\otimes \sigma_E^{\otimes 2}$.
%For the random coding, we use two independent actions on the two ports of the Choi states.
The parameters that determine the coding fidelity is $(K^4,N^2,M^4)$.
Then following the similar process it finally yields  
$\hat{Q}_{CA}(\Phi)\geq I(\omega_\C E,\Phi^{\otimes2})/4$ for any $\omega_\C E$.

\end{proof}

For the converse theorem, 
we need to define an entanglement generation task $E$ 
adopted to each model, which in general is the task 
to convert a certain type of input states into ebits, i.e. Bell states. 
The entanglement generation capacity satisfies $Q(\Phi)\leq E(\Phi)$,
and by showing any rate $\alpha$ for entanglement generation 
is smaller than $m(\rho,\Phi)$, 
together with the data-processing inequality,
it yields $Q(\Phi)\leq m(\rho,\Phi)$.
Note the entanglement generation is also called entanglement distribution sometimes. 

We specify four types of tasks in the following definition. 
For the Choi coding models, 
the input states are purified Choi states (see Sec.~\ref{sec:superc}) so that 
the input to a collection of channels  $ \Phi^{\otimes n}$ is 
the completely mixed state. 
Also let $|\omega_{(2)}\ket:=\frac{1}{d}\sum_i |i\ket |\omega_i\ket$ be a Bell state 
with $\{|\omega_i\ket\}$ as the Bell basis for $d^2$-dimensional space.
%Let $|\omega_{(3)}\ket:=\frac{1}{d}\sum_i |ii\ket |\omega_i\ket$ be a GHZ state. 
It is easy to see Bell states can be easily obtained from $|\omega_{(2)}\ket$. 
%and $|\omega_{(3)}\ket$.

\begin{definition}
(Entanglement generation capacities of a channel) 
Let $\Phi\in C(\C X, \C Y)$ be a channel, and an integer $k=\lfloor \alpha n \rfloor$ 
for all but finitely many positive integers $n$ and an achievable rate $\alpha \geq 0$, 
i) for the primary case, there exists a channel 
$\C D\in C(\C Y^{\otimes n}, \C Z^{\otimes k})$ and 
state $\xi\in \C X^{\otimes n}\otimes \C Z^{\otimes k}$ such that 
\be F(\omega^{\otimes k}, \C D \Phi^{\otimes n} \otimes \I^{\otimes k} (\xi))\geq 1- \epsilon, 
\label{eq:egf1} \ee
ii) for the EA case, there exists a channel 
$\C D\in C(\C W^{\otimes n}\otimes \C Y^{\otimes n}, \C Z^{\otimes k})$ and 
state $\xi \in  \C W^{\otimes n} \otimes \C X^{\otimes n}\otimes \C Z^{\otimes k}$ 
such that 
\be F(\omega^{\otimes k}, (\C D (\Phi\otimes \I)^{\otimes n})
\otimes \I^{\otimes k} (\xi))\geq 1- \epsilon, \label{eq:egf2}\ee
iii) for the Choi coding case, there exists a channel 
$\C D\in C(\C Y^{\otimes 2n}, \C Z^{\otimes 2k})$ and 
purified Choi state $\xi \in  \C X^{\otimes 2n}\otimes \C Z^{\otimes 2k}$ 
such that 
\be F(\omega_{(2)}^{\otimes k}, \C D \Phi^{\otimes 2n}
\otimes \I^{\otimes 2k} (\xi))\geq 1- \epsilon, \label{eq:egf3}\ee
iv) for the CA Choi coding case, there exists a channel 
$\C D\in C(\C W^{\otimes 2n}\otimes \C Y^{\otimes 2n}, \C Z^{\otimes 2k})$ and 
purified Choi state $\xi \in \C W^{\otimes 2n} \otimes \C X^{\otimes 2n}\otimes \C Z^{\otimes 2k}$ 
such that 
\be F(\omega_{(2)}^{\otimes k}, (\C D (\Phi\otimes \I)^{\otimes 2n})
\otimes \I^{\otimes 2k} (\xi))\geq 1- \epsilon, \label{eq:egf4}\ee
% there exists channel 
% $\C D\in C(\C Y^{\otimes n}\otimes \C W^{\otimes n}, \C Z^{\otimes m})$ and 
% purified Choi state $\xi$ such that the fidelity satisfies
% \be F(GHZ^{\otimes 2m}, (\C D (\Phi^{\otimes 2n}\otimes \I^{\otimes n})) \otimes \I^{\otimes 2m} (\xi))\geq 1- \epsilon\ee
for every choice of a positive real number $\epsilon\in[0,1]$,
and each entanglement generation capacity 
is defined as the supremum of all $\alpha$.
\end{definition}

For the proof, we also need the bounds 
 $\hat{Q}(\Phi)\leq \hat{E}(\Phi)$, 
$\hat{Q}_{CA}(\Phi)\leq \hat{E}_{CA}(\Phi)$. 
This is shown by using a Choi coding task with the completely depolarizing channel~\cite{Wat18} 
as the input to realize a corresponding entanglement generation task. 
This is the analog of $Q(\Phi)\leq E(\Phi)$, 
and $Q_{EA}(\Phi)\leq E_{EA}(\Phi)$.
Actually $Q(\Phi)= E(\Phi)$~\cite{Wat18},
and we conjecture that this equality can also be extended 
to the other three cases, 
but for the proof this is unnecessary.

\begin{proof}
(converse) The proof for the four models is similar.
i) For the usual entanglement generation, 
using the continuity of entropy the fidelity condition~(\ref{eq:egf1}) implies 
\bea S(\C D \Phi^{\otimes n} \otimes \I^{\otimes k} (\xi)) &\leq & 2\delta k+1, \\
     S(\C D \Phi^{\otimes n} (\rho)) &\geq& k-\delta k-1,
\eea
for $\delta\in (0,1)$ and 
$\rho\in \C D(\C X^{\otimes n})$ as a local state of $\xi$~\cite{Wat18}. 

% if a rate $\alpha$ is achievable, 
% then there must exist a state $\xi$ and channel $\C D$ so that 
% \be F(\omega^{\otimes k},(\C D \Phi^{\otimes n}\otimes \I^{\otimes k})(|\xi\ket))
% \geq 1-\epsilon \ee 
% and from the continuity of entropy, 

The coherent information satisfies the data-processing inequality 
$I_c(\rho,\Phi^{\otimes n})\geq I_c(\rho, \C D \Phi^{\otimes n})$~\cite{SN96},
so \be \frac{1}{n}I_c(\rho,\Phi^{\otimes n}) \geq \alpha (1-3\delta)  -\frac{3}{n}, \ee
this means $Q(\Phi)\leq E(\Phi)\leq I_{c_r}(\Phi)$. 

ii) For the EA entanglement generation, 
using the continuity of entropy the fidelity condition~(\ref{eq:egf2}) implies 
\bea S(\C D (\Phi\otimes\I)^{\otimes n} \otimes \I^{\otimes k} (\xi)) &\leq & 2\delta k+1, \\
     S(\C D (\Phi\otimes\I)^{\otimes n} (\rho)) &\geq& k-\delta k-1,
\eea
for $\rho\in \C D(\C X^{\otimes n}\otimes \C W^{\otimes n})$ as a local state of $\xi$. 

The mutual information satisfies the data-processing inequality 
$I(\rho,\Phi^{\otimes n})\geq I(\rho, \C D \Phi^{\otimes n})$ and also subadditivity~\cite{AC97},
so \be I(\rho,\Phi) \geq 2\alpha(1-3\delta/2)  -\frac{3}{n}, \ee
this means $Q_{EA}(\Phi)\leq E_{EA}(\Phi)\leq \max_\rho I(\rho,\Phi)/2$. 

iii) For the Choi coding entanglement generation, 
using the continuity of entropy the fidelity condition~(\ref{eq:egf3}) implies 
\bea S(\C D \Phi^{\otimes 2n} \otimes \I^{\otimes 2k} (\xi)) &\leq & 4\delta k+1, \\
     S(\C D \Phi^{\otimes 2n} (\rho)) &\geq& 2k-2\delta k-1,
\eea
for $\rho\in \C D(\C X^{\otimes 2n})$ as a local state of $\xi$. 

The Choi coherent information satisfies the data-processing inequality 
$I_c(\omega_\C E,\Phi^{\otimes 2n})\geq I_c(\omega_\C E, \C D \Phi^{\otimes 2n})$,
so \be \frac{1}{n}I_c(\omega_\C E,\Phi^{\otimes 2n}) \geq 2\alpha(1-3\delta) -\frac{3}{n}, \ee
this means $\hat{Q}(\Phi)\leq \hat{E}(\Phi)
\leq \max_{\omega_\C E} I_{c_r}(\omega_\C E,\Phi^{\otimes 2})$. 

iv) For the CA Choi coding entanglement generation, 
using the continuity of entropy the fidelity condition~(\ref{eq:egf4}) implies 
\bea S(\C D (\Phi\otimes\I)^{\otimes 2n} \otimes \I^{\otimes 2k} (\xi)) &\leq & 4\delta k+1, \\
     S(\C D (\Phi\otimes\I)^{\otimes 2n} (\rho)) &\geq& 2k-2\delta k-1,
\eea
for $\rho\in \C D(\C X^{\otimes 2n}\otimes \C W^{\otimes 2n})$ as a local state of $\xi$. 

The Choi mutual information satisfies the data-processing inequality 
$I(\omega_\C E,\Phi^{\otimes 2n})\geq I(\omega_\C E, \C D \Phi^{\otimes 2n})$ 
and also subadditivity,
so \be I(\omega_\C E,\Phi^{\otimes 2}) \geq 4\alpha(1-3\delta/2) -\frac{3}{n}, \ee
this means $\hat{Q}_{CA}(\Phi)\leq \hat{E}_{CA}(\Phi)\leq 
\max_{\omega_\C E} I(\omega_\C E,\Phi^{\otimes 2})/4$.
\end{proof}

\section{Quantum and classical CA Choi capacity}
\label{sec:qcchoi}

In this section, we study the relation between 
the quantum and classical CA Choi capacities. 
First of all, they can be viewed as the restricted version of 
the quantum and classical EA capacity, respectively, 
so we can use quantum teleportation and superdense coding to prove
$\hat{C}_{CA}(\Phi)=2\hat{Q}_{CA}(\Phi)$.

Similar with the EA setting, 
the noise-free control register could come from a pre-round of error correction.
A benefit of the assistance is to make the error-correction for some `bad' channels 
possible. 
For instance, the quantum capacity for the entanglement-breaking (EB) channel~\cite{HSR03} is zero,
but with EA, it is nonzero (except for the extremal case of replacing channel).
A seminal EA coding scheme is to use teleportation which effectively 
shuffle the input state $\rho$ into the noise-free register,
while the byproduct from the teleportation is corrected by sending 
the classical bits over the channel to the receiver. 
This scheme also extends to the CA Choi setting by using teleportation
for the two ports of Choi states. 

\begin{figure}[h!]
    \centering
    \includegraphics[width=0.45\textwidth]{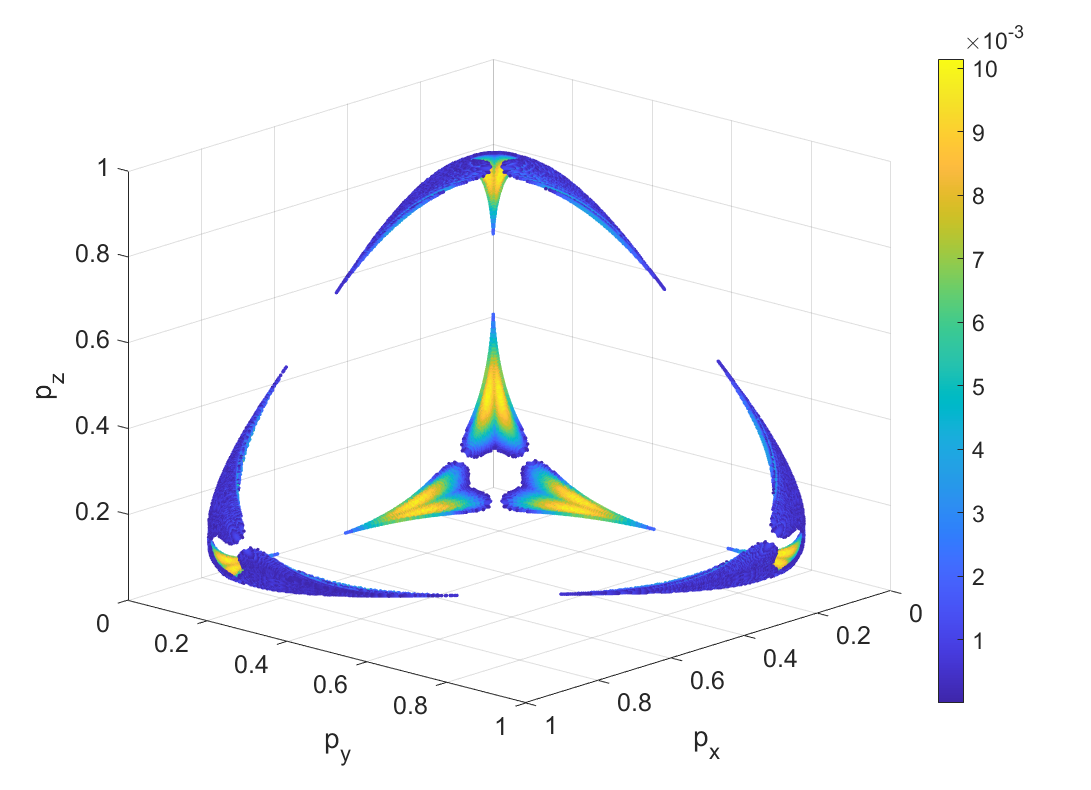}\vspace{0.0cm}
    \caption{The superadditive regions for the two-shot coherent information of the qubit Pauli channel.
    The colors of data points show the non-zero values of the difference between the two-shot coherent information
    $\max_\rho I_c(\rho,\C P^{\otimes 2})$
    and one-shot coherent information $\max_\rho I_c(\rho,\C P)$.}
    \label{fig:pauicapacity}
\end{figure}

Given a classical EA scheme, 
it can be used as a quantum EA scheme with twice of the rate 
by using quantum teleportation,
with the ebits as the additional entanglement assistance,
and the byproduct from teleportation sent via the classical EA scheme.
Given a quantum EA scheme, 
it can be used as a classical EA scheme with half of the rate
by using superdense coding, 
with the ebits as the additional entanglement assistance,
and the Bell measurement as a part of the decoder to extract the input bits. 
This proves the well-known result $C_{EA}(\Phi)=2Q_{EA}(\Phi)$~\cite{Wat18}.

Similarly, given a classical CA Choi protocol, 
we can use it to simulate a quantum CA Choi protocol with twice of the rate
by applying teleportation to each port of Choi states. 
The encoding remains as super-isometric, 
and the decoding involves the classical communication among the two ports
for the correction of Pauli byproducts. 

%we can use teleportation on each port to teleport B to C, and C to B.
% Then use the classical CA Choi protocol to error-correct the classical bits from 
% teleportation. 
% The decoder for the classical CA Choi protocol together with the Pauli byproduct 
% correction for teleportation form the whole decoder. 
%This shows $Q\geq C/2$.

Given a quantum CA Choi protocol, 
we can use it to simulate classical CA Choi protocol with half of the rate. 
We encode each bit value $x$ as an orthogonal purified Choi state 
$|\phi_{\Phi_x}\ket$, i.e., their dilated unitary operators $U_x$ are orthogonal,
then use the quantum CA Choi protocol to protect $|\phi_{\Phi_x}\ket$
followed by superdense coding to extract $x$.
In all, this proves $\hat{C}_{CA}(\Phi)=2\hat{Q}_{CA}(\Phi)$.

%We use the Choi state for each $U_{ABC}^x$, then we can do Bell measurement
%to distinguish them and obtain $x$. 
%The decoder in the quantum CA Choi protocol and Bell measurement form the whole decoder. 
%This shows $C\geq 2Q$.
%Then $C=2Q$.

This relation is also expected from the proofs for $\hat{Q}_{CA}$ and $\hat{C}_{CA}$.
For the direct theorem, the classical `version' of
the unitarily invariant ensemble 
is a random code construction according to the 
typicality of the input source, 
so that each code is also almost chosen uniformly~\cite{Wat18,Wil17}. 
The classical code dimension is the square root of the quantum case,
 and this will increase the classical capacity by a factor of two. 
For the converse theorem, 
the mutual information of classical maximally correlated bits (dephased ebits)
is half of the mutual information of ebits. 

%we also use the packing lemma,
%There is no need to consider a two-stage encoding
%so the code dimension is $K$ instead of $K^2$.

% CA Choi coding: the role of the control. 
% Two examples: 1 is use LCU and small control, noise-free, 2 is to correct EB noise, 
% we can use teleportation and this needs a large control.

%\newpage

\begin{figure}[t!]
    \centering
    \includegraphics[width=0.43\textwidth]{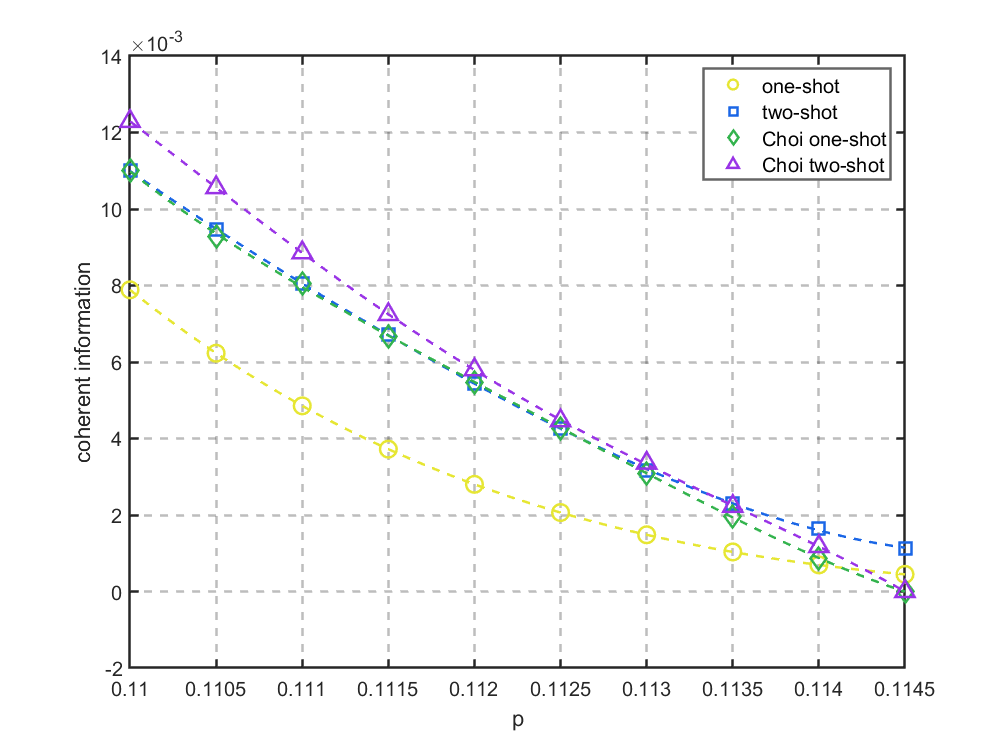}\vspace{0.0cm}
    \caption{The superadditivity for the two-shot coherent information of the dephrasure channel
    for $q=3p$.
    The four plots are for the one-shot $\max_\rho I_c(\rho,\C D)$ (circles),
    two-shot $\max_\rho I_c(\rho,\C D^{\otimes 2})$ (squares),
    one-shot $\max_{\omega_\C E} I_c(\omega_\C E,\C D^{\otimes 2})$ (diamond),
    and two-shot $\max_{\omega_\C E} I_c(\omega_\C E,\C D^{\otimes 4})$ (triangle).
    }
    \label{fig:dephrasure}
\end{figure}

\section{Examples of channels}
\label{sec:example}

Here we present examples of channel capacities including 
a numerical algorithm to compute them. 
We focus on the quantum capacities.
A large class of channels with additive coherent information are degradable channels~\cite{DS05},
such as the dephasing channels, erasure channels, 
and the qubit amplitude damping channels. 
Some non-degradable channels showing superadditivity are constructed recently~\cite{LLS18,LLS+23}. 

We developed an optimization algorithm to compute channel capacities.
The task is to find the optimal input $\rho$ that maximizes a coherent information.
However, the algorithm is not guaranteed to converge to the optimal value 
since the optimization is not convex. 
The algorithm is feasible only for small system sizes 
as the number of parameters of the input blows up exponentially. 

For Choi-type capacities, a sub-task is to generate random Choi states. 
There are many ways to achieve this, for instance, see Refs.~\cite{WS15,KNP21},
yet we find a very simple way to achieve this as a constraint 
in the GlobalSearch algorithm of Matlab.  
Namely, we first generate a random vector $\vec{r}$, 
with each value of it in the range $[-1,1]$,
and convert it into a non-negative semi-definite matrix $A$
using the Cholesky form~\cite{HJ91}.
A usual state $\rho$ is generated from $A$ with the trace-one constraint,
while a Choi state $\omega_{\C E}$ is generated from $A$ with the partial-trace constraint,
which is a collection of trace-one constraints.
These constraints can be naturally dealt with in the algorithm.

We numerically studied a few examples of channels.
A common channel is the qubit Pauli channel $\C P$ with Kraus operators $\sqrt{p_i}\sigma_i$, 
which, however, is non-degradable~\cite{DSS98,SS07,FW08,FKG20,SAB22}.
We computed the two-shot coherent information,
and confirmed the recent finding of Ref.\cite{SAB22}.
There is a superadditive region for small values of error probabilities
but a large no-error event $p_0$, see Fig.~\ref{fig:pauicapacity}.
There are also three regions for small $p_0$, 
which are physically equivalent to the large $p_0$ region
by acting a Pauli operator after $\C P$.
For each region, 
the optimal input states are rank-two maximally correlated mixed states, 
e.g., $(|00\ket \bra 00|+|11\ket \bra 11|)/2$,
and its equivalents under local Pauli operators and joint qubit Clifford operators,
leading to six sub-regions for a superadditive region~\cite{SAB22}.
Also we observe that the optimal input states 
do not depend on the values of probabilities, 
and they are Choi states,
so for this setting there is no separation 
between the Choi coherent information and the usual coherent information.

We also numerically studied the dephrasure channel~\cite{LLS18} defined as 
$\C D(\rho)=(1-q)[(1-p)\rho+p \sigma_z \rho \sigma_z]+q |e\ket \bra e|$
for $p,q\in [0,1]$ and $|e\ket$ is orthogonal to the input space,
and confirmed the two-shot superadditivity phenomenon.
The Choi coherent information is also superadditive,
as shown in the Fig.~\ref{fig:dephrasure}.
Different from the Pauli channel,
the optimal input depends on the channel parameters $(p,q)$.
We also find differences among the usual coherent information 
and Choi coherent information.
For instance, 
the optimal input for the two-shot coherent information (squared data points)
is not a Choi state,
hence it is larger than the Choi coherent information (diamond data points). 
Also we find the one-shot coherent information (circled data points)
can be larger than Choi coherent information for some values of parameters 
($p=0.1145$ in the figure).
Therefore, this is an example that shows a separation 
between the Choi coherent information and the usual coherent information.

\end{spacing}
\bibliography{ext}{}

\begin{thebibliography}{10}
\expandafter\ifx\csname url\endcsname\relax
  \def\url#1{\texttt{#1}}\fi
\expandafter\ifx\csname urlprefix\endcsname\relax\def\urlprefix{URL }\fi
\expandafter\ifx\csname href\endcsname\relax
  \def\href#1#2{#2} \def\path#1{#1}\fi

\bibitem{NC00}
M.~A. Nielsen, I.~L. Chuang, Quantum Computation and Quantum Information,
  Cambridge University Press, Cambridge U.K., 2000.

\bibitem{LB13}
D.~Lidar, T.~A. Brun (Eds.), Quantum error correction, Cambridge University
  Press, 2013.

\bibitem{Wil17}
M.~Wilde, Quantum Information Theory, Cambridge University Press, 2017.

\bibitem{Wat18}
J.~Watrous, {The Theory of Quantum Information}, Cambridge University Press,
  2018.

\bibitem{SN96}
B.~Schumacher, M.~A. Nielsen, Quantum data processing and error correction,
  Phys. Rev. A 54 (1996) 2629--2635.

\bibitem{Llo97}
S.~Lloyd, Capacity of the noisy quantum channel, Phys. Rev. A 55 (1997)
  1613--1622.

\bibitem{BNS98}
H.~Barnum, M.~A. Nielsen, B.~Schumacher, Information transmission through a
  noisy quantum channel, Phys. Rev. A 57 (1998) 4153--4175.

\bibitem{SY08}
G.~Smith, J.~Yard, Quantum communication with zero-capacity channels, Science
  321 (2008) 1812.

\bibitem{Hol99}
A.~S. Holevo, Quantum coding theorems, Russian Math. Surveys 53 (1999)
  1295–1331.

\bibitem{Has09}
M.~B. Hastings, Superadditivity of communication capacity using entangled
  inputs, Nat. Phys. 5 (2009) 255.

\bibitem{Dev05}
I.~Devetak, {The private classical capacity and quantum capacity of a quantum
  channel}, IEEE Trans. Inf. Theory 51~(1) (2005) 44--55.

\bibitem{LLS+14}
D.~Leung, K.~Li, G.~Smith, J.~A. Smolin, Maximal privacy without coherence,
  Phys. Rev. Lett. 113 (2014) 030502.

\bibitem{DHW04}
I.~Devetak, A.~W. Harrow, A.~Winter, A family of quantum protocols, Phys. Rev.
  Lett. 93 (2004) 230504.

\bibitem{ADHW09}
A.~Abeyesinghe, I.~Devetak, P.~Hayden, A.~Winter, The mother of all protocols:
  Restructuring quantum information's family tree, Proc. R. Soc. A 465 (2009)
  2537--2563.

\bibitem{CLS17}
A.~Cross, K.~Li, G.~Smith, Uniform additivity in classical and quantum
  information, Phys. Rev. Lett. 118 (2017) 040501.

\bibitem{Jam72}
A.~Jamio{\l}kowski, Linear transformations which preserve trace and positive
  semidefiniteness of operators, Rep. Math. Phys. 3 (1972) 275.

\bibitem{Cho75}
M.-D. Choi, Completely positive linear maps on complex matrices, Linear Algebra
  Appl. 10 (1975) 285--290.

\bibitem{CDP08}
G.~Chiribella, G.~M. D'Ariano, P.~Perinotti, Quantum circuit architecture,
  Phys. Rev. Lett. 101 (2008) 060401.

\bibitem{CG19}
E.~Chitambar, G.~Gour, Quantum resource theories, Rev. Mod. Phys. 91 (2019)
  025001.

\bibitem{W22_qvn}
D.-S. Wang, {A prototype of quantum von Neumann architecture}, Commun. Theor.
  Phys. 74 (2022) 095103.

\bibitem{W24_qvn}
D.-S. Wang, {A family of quantum von Neumann architecture}, Chin. Phys. B 33
  (2024) 080302.

\bibitem{BKN00}
H.~Barnum, E.~Knill, M.~Nielsen, {On quantum fidelities and channel
  capacities}, IEEE Trans. Inf. Theory 46~(4) (2000) 1317--1329.

\bibitem{HHW+08}
P.~Hayden, M.~Horodecki, A.~Winter, J.~Yard, A decoupling approach to the
  quantum capacity, Open Syst. Inf. Dyn. 15 (2008) 7--19.

\bibitem{Kle07}
R.~Klesse, Approximate quantum error correction, random codes, and quantum
  channel capacity, Phys. Rev. A 75 (2007) 062315.

\bibitem{BSS+99}
C.~H. Bennett, P.~W. Shor, J.~A. Smolin, A.~V. Thapliyal, Entanglement-assisted
  classical capacity of noisy quantum channels, Phys. Rev. Lett. 83 (1999)
  3081--3084.

\bibitem{BSST02}
C.~H. Bennett, P.~W. Shor, J.~A. Smolin, A.~V. Thapliyal, Entanglement-assisted
  capacity of a quantum channel and the reverse shannon theorem, IEEE Trans.
  Inf. Theory 48 (2002) 2637.

\bibitem{BDH+14}
C.~H. Bennett, I.~Devetak, A.~W. Harrow, P.~W. Shor, A.~Winter, The quantum
  reverse shannon theorem and resource tradeoffs for simulating quantum
  channels, IEEE Trans. Inf. Theory 60 (2014) 2926.

\bibitem{AC97}
C.~Adami, N.~J. Cerf, {von Neumann capacity of noisy quantum channels}, Phys.
  Rev. A 56 (1997) 3470--3483.

\bibitem{MW14}
W.~Matthews, S.~Wehner, Finite blocklength converse bounds for quantum
  channels, IEEE Trans.Info.Theor. 60 (2014) 7317.

\bibitem{LM15}
D.~Leung, W.~Matthews, On the power of ppt-preserving and non-signalling codes,
  IEEE Trans.Info.Theor. 61 (2015) 4486.

\bibitem{WLWL24}
D.-S. Wang, Y.-D. Liu, Y.-J. Wang, S.~Luo, Quantum resource theory of coding
  for error correction, Phys. Rev. A 110 (2024) 032413.

\bibitem{Sch96}
B.~Schumacher, Sending entanglement through noisy quantum channels, Phys. Rev.
  A 54 (1996) 2614--2628.

\bibitem{KL97}
E.~Knill, R.~Laflamme, Theory of quantum error-correcting codes, Phys. Rev. A
  55 (1997) 900--911.

\bibitem{Got98}
D.~Gottesman, Theory of fault-tolerant quantum computation, Phys. Rev. A 57
  (1998) 127--137.

\bibitem{SS07}
G.~Smith, J.~A. Smolin, Degenerate quantum codes for pauli channels, Phys. Rev.
  Lett. 98 (2007) 030501.

\bibitem{Bow02}
G.~Bowen, Entanglement required in achieving entanglement-assisted channel
  capacities, Phys. Rev. A 66 (2002) 052313.

\bibitem{DS05}
I.~Devetak, P.~W. Shor, The capacity of a quantum channel for simultaneous
  transmission of classical and quantum information, Commun. Math. Phys. 256
  (2005) 287–303.

\bibitem{HHH00}
M.~Horodecki, P.~Horodecki, R.~Horodecki, Unified approach to quantum
  capacities: Towards quantum noisy coding theorem, Phys. Rev. Lett. 85 (2000)
  433--436.

\bibitem{LLS18}
F.~Leditzky, D.~Leung, G.~Smith, Dephrasure channel and superadditivity of
  coherent information, Phys. Rev. Lett. 121 (2018) 160501.

\bibitem{LLS+23}
F.~Leditzky, D.~Leung, V.~Siddhu, G.~Smith, J.~A. Smolin, Generic nonadditivity
  of quantum capacity in simple channels, Phys. Rev. Lett. 130 (2023) 200801.

\bibitem{W20_choi}
D.-S. Wang, Choi states, symmetry-based quantum gate teleportation, and
  stored-program quantum computing, Phys. Rev. A 101 (2020) 052311.

\bibitem{BCC+15}
D.~W. Berry, A.~M. Childs, R.~Cleve, R.~Kothari, R.~D. Somma, Simulating
  hamiltonian dynamics with a truncated taylor series, Phys. Rev. Lett. 114
  (2015) 090502.

\bibitem{BCC+14}
D.~W. Berry, A.~M. Childs, R.~Cleve, R.~Kothari, R.~D. Somma, Exponential
  improvement in precision for simulating sparse hamiltonians, in: Proc. 46th
  ACM Symposium on Theory of Computing, 2014, p. 283.

\bibitem{WW23}
K.~Wang, D.-S. Wang, Quantum circuit simulation of superchannels, New J. Phys.
  25~(4) (2023) 043013.

\bibitem{LWLW23}
Y.-T. Liu, K.~Wang, Y.-D. Liu, D.-S. Wang, {A Survey of Universal Quantum von
  Neumann Architecture}, Entropy 25~(8) (2023) 1187.

\bibitem{GC99}
D.~Gottesman, I.~L. Chuang, Demonstrating the viability of universal quantum
  computation using teleportation and single-qubit operations, Nature
  402~(6760) (1999) 390--393.

\bibitem{BZ06}
I.~Bengtsson, K.~\.{Z}yczkowski, Geometry of Quantum States, Cambridge
  University Press, Cambridge U.K., 2006.

\bibitem{NC97}
M.~A. Nielsen, I.~L. Chuang, Programmable quantum gate arrays, Phys. Rev. Lett.
  79 (1997) 321--324.

\bibitem{YRC20}
Y.~Yang, R.~Renner, G.~Chiribella, Optimal universal programming of unitary
  gates, Phys. Rev. Lett. 125 (2020) 210501.

\bibitem{AFC14}
M.~Araujo, A.~Feix, F.~Costa, C.~Brukner, Quantum circuits cannot control
  unknown operations, New J. Phys. 16 (2014) 093026.

\bibitem{CAPV13}
G.~Chiribella, G.~M. D'Ariano, P.~Perinotti, B.~Valiron, Quantum computations
  without definite causal structure, Phys. Rev. A 88 (2013) 022318.

\bibitem{SW02}
B.~Schumacher, M.~D. Westmoreland, Approximate quantum error correction,
  Quantum Infor. Processing 1 (2002) 512.

\bibitem{Weyl46}
H.~Weyl, The Classical Groups, princeton University Press, Princeton, NJ, 1946.

\bibitem{HSR03}
M.~Horodecki, P.~Shor, M.~B. Ruskai, Entanglement breaking channels, Rev. Math.
  Phys. 15 (2003) 629--641.

\bibitem{WS15}
D.-S. Wang, B.~C. Sanders, {Quantum circuit design for accurate simulation of
  qudit channels}, New J. Phys. 14~(3) (2015) 033016.

\bibitem{KNP21}
R.~Kukulski, I.~Nechita, l.~Pawela, Z.~Pucha\l{}a, K.~\.{Z}yczkowski,
  Generating random quantum channels, J. Math. Phys. 62 (2021) 062201.

\bibitem{HJ91}
R.~A. Horn, C.~R. Johnson, Topics in Matrix Analysis, Cambridge University
  Press, Cambridge, U.K., 1991.

\bibitem{DSS98}
D.~P. DiVincenzo, P.~W. Shor, J.~A. Smolin, Quantum-channel capacity of very
  noisy channels, Phys. Rev. A 57 (1998) 830--839.

\bibitem{FW08}
J.~Fern, K.~B. Whaley, Lower bounds on the nonzero capacity of pauli channels,
  Phys. Rev. A 78 (2008) 062335.

\bibitem{FKG20}
M.~Fanizza, F.~Kianvash, V.~Giovannetti, Quantum flags and new bounds on the
  quantum capacity of the depolarizing channel, Phys. Rev. Lett. 125 (2020)
  020503.

\bibitem{SAB22}
G.~L. Sidhardh, M.~Alimuddin, M.~Banik, Exploring superadditivity of coherent
  information of noisy quantum channels through genetic algorithms, Phys. Rev.
  A 106 (2022) 012432.

\end{thebibliography}
\bibliographystyle{elsarticle-num}

\end{document}